\documentclass[aps, prd, superscriptaddress, nofootinbib, preprintnumbers]{revtex4}
\usepackage[dvipdfmx]{graphicx}
\usepackage{graphicx,bm,color}
\usepackage{subfigure}
\usepackage{amsmath}
\usepackage{amssymb}
\usepackage{amsfonts}
\usepackage{cases}
\usepackage{cancel}
\usepackage{hyperref}
\usepackage{ulem}

\usepackage{here}
\usepackage{tikz}
\usetikzlibrary{matrix}

\begin{document}

\title{
Restriction on 
the form of quark anomalous magnetic moment
from lattice QCD results
}

\author{Mamiya Kawaguchi}\thanks{{\tt mamiya@ucas.ac.cn}} 

\author{Mei Huang}\thanks{{\tt huangmei@ucas.ac.cn}}
\affiliation{School of Nuclear Science and Technology, University of Chinese Academy of Sciences, Beijing 100049, China}

\begin{abstract}
The quark anomalous magnetic moment (AMM) is dynamically generated through the spontaneous chiral symmetry breaking. It has been revealed that even though its exact form is still unknown, the quark AMM is essential to explore quark matter properties and QCD phase structure under external magnetic fields. In this study, we take three different forms of the quark AMM and investigate its influence on the chiral phase transition under magnetic field. In general, a negative quark AMM plays the role as magnetic catalyzer and a positive quark AMM plays the role of magnetic inhibition. It is found that a constant quark AMM drives an unexpected 1st order chiral phase transition;  a quark AMM proportional to the chiral condensate gives a flip of the sign on the chiral condensate; and a quark AMM proportional to the square of chiral condensate can produce results of chiral condensate as functions of the temperature and the magnetic field in good agreement with the lattice result.
\end{abstract}

\maketitle
\section{Introduction}

Exploring the properties of the magnetized hot/dense quark mater is a subject of great interest in the high energy nuclear physics relevant to the neutron stars 
and the quark-gluon plasma created in non-central relativistic heavy ion collisions. 
In such a thermomagnetic system,
striking phenomena emerge and are expected to bring us a new aspect of the nonperturbative feature of QCD.
In particular,  the influence of an external magnetic field on
the QCD phase transition is one of important phenomena 
in understanding of the quark matter under extreme conditions. 

The nonperturbative phenomenon of the dynamical chiral symmetry breaking is crucially affected by  external magnetic fields in hot/dense QCD matter. 
At low temperatures when the chiral symmetry is broken, the magnetic field enhances the chiral condensate, which acts as a catalyzer for the spontaneous chiral symmetry breaking. This magnetic catalysis (MC) behavior has been observed in the Nambu-Jona-Lasinio (NJL) model~\cite{Klevansky:1989vi,Klimenko:1990rh,Gusynin:1995nb}, in effective model approaches like the chiral perturbation theory \cite{Andersen:2012zc},
as well as 
in the lattice QCD simulations~\cite{Bali:2011qj,Bali:2012zg,Bali:2013esa}, 

The effective model analysis provides the clear interpretation for the MC: the magnetic dimensional reduction induces the enhancement of the chiral symmetry breaking and then the MC is realized in the vacuum. In contrast to the case at low temperatures, 
the inverse magnetic catalysis (IMC) arises around the chiral phase transition temperature, which has been observed by the lattice QCD simulation  ~\cite{Bali:2011qj,Bali:2012zg,Bali:2013esa}:
the magnetic field promotes the chiral symmetry restoration and reduces the critical temperature of the chiral phase transition. As it is expected, all effective chiral models including the conventional NJL model display the magnetic catalysis at low temperatures as well as at high temperatures~\cite{Klevansky:1989vi,Klimenko:1990rh,Gusynin:1995nb}. 
In order to explain the IMC behavior, extra mechanism has to be taken into account, for example, the neutral pion fluctuation \cite{Fukushima:2012kc}, the chirality imbalance \cite{Chao:2013qpa} and the running coupling with magnetic field ~\cite{Ferrer:2014qka}.


Recently, the anomalous magnetic moment (AMM) of quarks attracts much interest to gain a new insight of QCD matter under magnetic field. In the perturbative framework of the massless QCD,  it is prohibited to provide the AMM contribution for quarks due to the presence of the chiral symmetry. However, when the chiral symmetry is dynamically broken at the low energy regime of QCD, the quarks would possess the AMM terms. It has been shown in~\cite{Chang:2010hb} that the spontaneous chiral symmetry breaking dynamically generates the AMM even for massless quarks in the absence of the external magnetic field. The magnetic field effect has been also studied based on the NJL model with the tensor interactions \cite{Ferrer:2013noa,Mao:2018jdo,Mei:2020jzn,Xu:2020yag,Lin:2022ied},  in which the strong magnetic field induces the dynamical generation of the AMM associated with the emergence of the spin condensate. It has been observed in ~\cite{Ferrer:2014qka} that the existence of the dynamical AMM leads to the IMC behavior in the strong magnetic field region even though only the lowest Landau level is taken into account in quarks, and it was observed in \cite{Lin:2022ied} that with AMM, the behavior of neutral pion mass and charged pion mass under magnetic field qualitatively in agreement with lattice result \cite{Ding:2020jui}. Recently, it was shown in \cite{Xu:2022hql} that the AMM of quarks contributes 40$\%$ to the $\bar{\Lambda}-\Lambda$ polarization splitting.

Although the AMM of quarks would be expected to be a key ingredient for investigating the thermomagnetic QCD matter, its expression remains obscure at the low energy regime of QCD.
In the vacuum, conventionally the AMM of quarks is treated as a constant value, and 
the influence of the constant AMM on the chiral phase transition has been investigated. 
In the NJL model~\cite{Fayazbakhsh:2014mca,Chaudhuri:2019lbw} and the PNJL model~\cite{Mei:2020jzn}, these studies have shown that the constant AMM contribution drastically affects the chiral phase transition  and triggers 
an unexpected phase transition contradicting with the lattice observation 
(this AMM effect on the phase transition will be discussed in the later section). 
Hence, the constant AMM would be a improper form for QCD matter under magnetic field.
The alternative expression of the AMM, which involves the behavior proportional to the chiral condensate or the square of chiral condensate, has been also suggested in the NJL model 
to evaluate the magnetic effect on the meson masses and the magnetic susceptibility in \cite{Xu:2020yag,Lin:2022ied}. However,
the details on how the form of the quark AMM will affect the chiral phase transition has not been addressed in  \cite{Xu:2020yag,Lin:2022ied}.
Given these facts, there is still room left for constraining the effective form of the quark AMM at the low-energy QCD 
while taking account of the chiral phase transition under magnetic fields.

In the present paper, we explore the effective form of the quark AMM which can properly describe the thermomagnetic QCD matter. Considering the uncertainty of the explicit expression for the quark AMM, we make assumptions about the effective form of the quark AMM. 
Firstly, we regard the AMM as a constant value for the sake of simplicity. Next we suppose that the AMM depends on the scalar meson field serving as the chiral order parameter.
In the later case, we deduce the effective form of the AMM from the above findings in
\cite{Chang:2010hb,Fayazbakhsh:2014mca,Chaudhuri:2019lbw} and the recent study 
about the magnetic effect on the electron AMM within the QED framework \cite{Lin:2021bqv}.
With the effective AMM forms, we evaluate the influence of the quark AMM on the chiral phase transition based on the two-flavor NJL model and consider the practicable expression of the effective AMM form  
by comparing  the NJL results with the lattice QCD results. Through the comparison, we also show the intrinsic temperature dependence of the quark AMM.
The phenomenological implications of the effective  
AMM are also discussed.

This paper is organized as follows. In Sec.~\ref{sec1}, we introduce the NJL model taking into account the effective interaction of the quark AMM,
and
assume several forms of the quark AMM.
Then we evaluate the influence of the quark AMM on the chiral phase transition under the constant magnetic field and compare the NJL results with the lattice QCD observations in Sec.~\ref{sec2}.
Summary and discussion are given in  Sec.~\ref{sec3}.

\section{
low energy description of Quark AMM  
}
\label{sec1}

The regular NJL model consists of four-quark point interactions, which can describe the chiral symmetry breaking in the vacuum and the symmetry restoration at finite-temperatures. As a minimal extension, we work on the NJL model involving the AMM term in order to explore the effective form of the quark AMM and its influence on the chiral phase transition.
Note that we suppose that the chiral condensate and the quark AMM take the isospin symmetric form to simply consider the correlation between the spontaneous chiral symmetry breaking and the AMM contribution. 

In this section,  we firstly briefly introduce the NJL framework. 
Then, we show 
several forms of the quark AMM 
in the NJL-model description.

\subsection{Nambu--Jona-Lasinio model with quark AMM term}
The two-flavor NJL model involving the AMM term is written as
\begin{eqnarray}
{\cal L}_{\rm NJL}&=&
\bar \psi \left(i\gamma^\mu D_\mu -{\bm m}+\frac{1}{2}\kappa_f q_f F_{\mu\nu} \sigma^{\mu\nu} \right)\psi
+G_S\left\{
(\bar \psi \psi)^2 +(\bar \psi i\gamma_5\vec \tau \psi)^2
\right\}
\label{NJL}
\end{eqnarray}
where 
$\psi$ denotes the two-flavor quark field,  ${\psi}=(u,d)^T$;
${\bm m}$ represents a matrix of the current quark mass ${\bm m}={\rm diag}(m_u,m_d)$ 
and we take the isospin symmetric limit $m_u=m_d=m_0$, $\sigma_{\mu\nu}=\frac{i}{2}[\gamma_\mu, \gamma_\nu]$; the field strength of the gauge field $A_\mu$ is given by  $F_{\mu\nu}=\partial_\mu A_\nu -\partial_\nu A_\mu$;
$G_S$ is the coupling constant in the scalar channel of the four-quark interaction term.
 The covariant derivative acting on the quark field is represented as $D_\mu=\partial_\mu -iq_fA_\mu$ with the electric charge for quark flavor $q_u=+2e/3$ and $q_d=-e/3$.
In this study, we consider a constant magnetic field along the z-direction in the four-dimensional spacetime, which
is embedded in the gauge field $A_\mu=(0,-By/2,Bx/2,0)$. 
The interaction form of the quark AMM is introduced as $\bar \psi \kappa_f q_f F_{\mu\nu} \sigma^{\mu\nu}\psi$, with the quark AMM $\kappa_f$ 
and its expression will be given latter.



Introducing the auxiliary scalar, pseodoscalar fields $\sigma\sim \bar \psi \psi$ and $\vec{\pi} \sim \bar \psi i\gamma_5\vec{\tau}\psi$, and taking the mean field approximation $\sigma=-2G_S\bar \psi \psi$, and $\pi=0$,
one can derive the bosonized NJL Lagrangian as
\begin{eqnarray}
{\cal L}_{\rm mean}=
\bar \psi \left(i\gamma^\mu D_\mu -M+\frac{1}{2}\kappa_f q_f F_{\mu\nu} \sigma^{\mu\nu} \right)\psi
-\frac{1}{4G_S}\sigma^2,
\end{eqnarray}
where $M=m_0+\sigma$. 
By integrating out the quark field in the generating functional for the mean-field Lagrangian ${\cal L}_{\rm mean}$, 
one obtains the effective potential of the NJL model, 
\begin{eqnarray}
V_{\rm eff} (m_0,\sigma, T, eB)=
\frac{\sigma^2}{4G_S}
-N_c\sum_{q_f=q_u,q_d}
|q_f B|
\sum_{l=0}^\infty
\sum_{s=\pm1}
\int \frac{dp_3}{4\pi^2}
\left\{
E_{q_f}^{(l,s)}f_{\Lambda,B}^{(l,s)}
+2T\ln\left(
1+e^{- E_{q_f}^{(l,s)}/T  }
\right)
\right\}
\label{V_eff}
\end{eqnarray}
where 
$N_c$ denotes the number of colors,
$l$ labels the Landau level, 
$s=\pm 1$ represents the spin-up/down of the quarks
and  $E_{q_f}^{(l,s)}$  is
the energy dispersion relation of up- and down-quarks in the external magnetic field and with the quark AMM,
\begin{eqnarray}
E_{q_f}^{(l,s)}=
\sqrt{
p_3^2+
\left[
\left\{
|q_f B|(
2l+1 -s \xi_f
)+M^2
\right\}^{1/2}
-s \kappa_fq_f B
\right]^2
}
.
\end{eqnarray}
with
$p_3$ denoting the third component in the momentum space and 
$\xi_f={\rm sgn}(q_fB)$.
The vacuum part of the effective potential involves
an ultraviolet divergence.
To regularize the divergence, 
we use a a smooth regularization scheme
by inserting 
the regulator function $f_{\Lambda,B}^{(l,s)}$ into the effective potential in Eq.~(\ref{V_eff}).
In this work, we choose the Lorenztian form factor as follows,
\begin{eqnarray} 
f_{\Lambda,B}^{(l,s)}
=
\frac{\Lambda^{10}}{\Lambda^{10}+(\sqrt{p_3^2 +|q_f B|(2l+1-s\xi_f )})^{10}
},
\label{regulator}
\end{eqnarray}
with
$\Lambda$ being the ultraviolet momentum cutoff.

The expectation value of the auxiliary scalar field
$\bar \sigma=\langle \sigma \rangle$ can be determined from the stationary condition for the effective potential, 
\begin{eqnarray}
\frac{\partial V_{\rm eff}(\sigma) }{\partial \sigma}\Biggl|_{\sigma=\bar \sigma}=0.
\end{eqnarray}
Indeed, $\bar \sigma$ plays a role of the order parameter for the spontaneous chiral symmetry breaking, which is called the chiral condensate, and it gives the dynamical quark mass generated by the spontaneous chiral symmetry breaking $\langle M\rangle=m_0+\bar{\sigma}$. 
Using the chiral condensate, one can obtain 
the quark condensate,
\begin{eqnarray}
\langle \bar \psi \psi \rangle =
\frac{\partial V_{\rm eff}}{\partial m_0}\Biggl|_{\sigma=\bar \sigma}.
\end{eqnarray}

\subsection{The form of the quark AMM}
As was shown in the NJL Lagrangian Eq.~(\ref{NJL}), the AMM interaction term for quarks can be expressed as 
\begin{eqnarray}
{\cal L}_{\rm int}^{\rm (AMM)}=
 \frac{1}{2}\kappa_f q_f \bar \psi F_{\mu\nu} \sigma^{\mu\nu} \psi.
 \label{AMM_int}
\end{eqnarray}
As mentioned in the Introduction, in the Bethe-Salpeter approach for the quark-photon vertex without the external magnetic field correction, it has been argued in \cite{Chang:2010hb} that 
the AMM is dynamically generated from the spontaneous chiral symmetry breaking. 
 As a consequence, the AMM term vanishes at the chiral limit when the chiral symmetry is completely restored. Though the explicit form of $\kappa_{f}$ remains unclear, the quarks dynamically acquire an AMM term, thus naively one can expect that the quark AMM $\kappa_{u,d}$ should depend on the chiral condensate, which describes the chiral symmetry breaking term.
Furthermore, when we consider quark matter under external magnetic fields,  the magnetic field correction would make the AMM form more complicated.

The external magnetic field induces the dynamical generation of AMM which can be called the magnetic-dependent AMM. In the NJL model with tensor interactions \cite{Ferrer:2013noa,Mao:2018jdo,Xu:2020yag,Lin:2022ied}, 
the spin polarization condensate $\langle \bar \psi \sigma^{\mu\nu} \psi\rangle$ is generated and it is proportional to the chiral condensate under the strong magnetic field region, which is similar to the quark AMM.
This indicates that the chiral condensate would enter the AMM form 
by taking into account the external magnetic field dependence on the vacuum. We can assume the magnetic-dependent AMM proportional to the chiral condensate  $\kappa_{u,d}\sim \sigma$, as one possible form of the quark AMM. 

In addition, we also raise an alternative possibility for the expression of the magnetic-dependent AMM. In the QED framework, it is found in \cite{Lin:2021bqv} that the external magnetic field actually contributes to   the AMM for the electron, which is proportional to the electron mass squared with magnetic corrections. From this fact, we naively extend the QED framework to the QCD case and promote the electron mass to the dynamical quark mass. Then we can assume that the magnetic-dependent AMM  takes the form of $\kappa_{u,d}\sim M^2\sim \sigma^2$.




Given the above facts, we consider the following three cases separately
to find out the effective form of the quark AMM in the low energy regime of QCD
\footnote{
Although other expression of $\kappa_f$ would be conceivable, in this study we focus on only the three cases for simplicity.
},  
\begin{itemize}
    \item[(a)] $\kappa_{u,d}={\rm const.}$;
    \item[(b)] $\kappa_{u,d}=v \sigma$; 
    \item[(c)] $\kappa_{u,d}=\bar v \sigma^2$,
\end{itemize}
where $v$ and $\bar v$ are free parameters in the low energy effective model.

In the case of (a), we assume that the AMM $\kappa_{f}$ takes the constant value for sake of simplicity.
Indeed, the constant AMM for quarks is evaluated from the proton and neutron magnetic moment by using the constituent quark model \cite{Fayazbakhsh:2014mca}:
$\kappa_u= 0.29016\,{\rm GeV}^{-1}$ and $\kappa_d= 0.35986\,{\rm GeV}^{-1}$.
In the case of (b) and (c), we give an assumption that $\kappa_f$ takes 
the magnetic-dependent forms in the magnetized vacuum. 
These forms actually vanish after the chiral restoration. This implies that the magnetic-dependent AMM may be dynamically generated via the spontaneous chiral symmetry breaking.

Note that the NJL model including the magnetic-dependent AMM has been discussed to study the AMM contribution on the meson masses and the magnetic susceptibility \cite{Xu:2020yag,Lin:2022ied}.


In this paper, we investigate the influence of the quark AMM on the chiral phase transition in order to find out the effective form of $\kappa_f$ through the comparison between the NJL's estimates and the recent lattice QCD observations.
\section{
Influence of quark AMM on chiral phase transition
}
\label{sec2}
In this section, we numerically evaluate the AMM dependence on the chiral condensate (quark condensate) based on the NJL model. 
Comparing the (subtracted) quark condensate with the recent lattice QCD data \cite{Ding:2020hxw, Ding:2022tqn}
\footnote{In \cite{Ding:2020hxw, Ding:2022tqn}, 
the pion mass at $T=0$ and $eB=0$ is taken as $m_\pi\simeq 220$ MeV somewhat deviated from the physical value 
and the continuum limit is not taken, 
 which are qualitatively consistent with the other lattice observation \cite{Bali:2012zg} (the physical pion mass and the continuum limit are taken into account in \cite{Bali:2012zg}). 
}, we restrict the effective form of the quark AMM at the low energy dynamics of QCD under the magnetic field.

The model parameters are fixed as 
$\Lambda = 681.38\,{\rm GeV}$,
$G_S\Lambda^2= 1.860$,
$m_0=4.552\,{\rm MeV}$
\cite{Avancini:2019wed}.
With this parameter set, the following physical observables are provided at $T=0$, 
the dynamical mass $M=286.19$ MeV,
the pion mass $m_\pi=138$ MeV, and
the pion decay constant $f_\pi=92.4$ MeV.

The present NJL model experiences a chiral crossover at $eB=0$. 
The pseudo-critical temperature of the chiral crossover 
is defined by 
$d^2\bar \sigma(T,eB=0)/d T^2\bigl|_{T=T_{\rm pc}}=0$.
We find $T_{\rm pc}^{\rm ( NJL)}(eB=0) \simeq167$ MeV, which is close to the  Lattice QCD simulation result  $T_{\rm pc}^{\rm (lat.)}(eB=0)\simeq 170$ MeV in \cite{Ding:2022tqn}.

Below, we investigate the influence of the AMM on the chiral condensate by taking the following three forms of quark AMM separately:
(a) $\kappa_{u,d}={\rm const.}$,
(b) $\kappa_{u,d}=v \sigma$ and
(c) $\kappa_{u,d}=\bar v \sigma^2$.



\subsection{Constant AMM: $\kappa_{u,d}={\rm const.}$}

\begin{figure}[H] 
\begin{tabular}{cc}
\begin{minipage}{0.5\hsize}
\begin{center}
    \includegraphics[width=8cm]{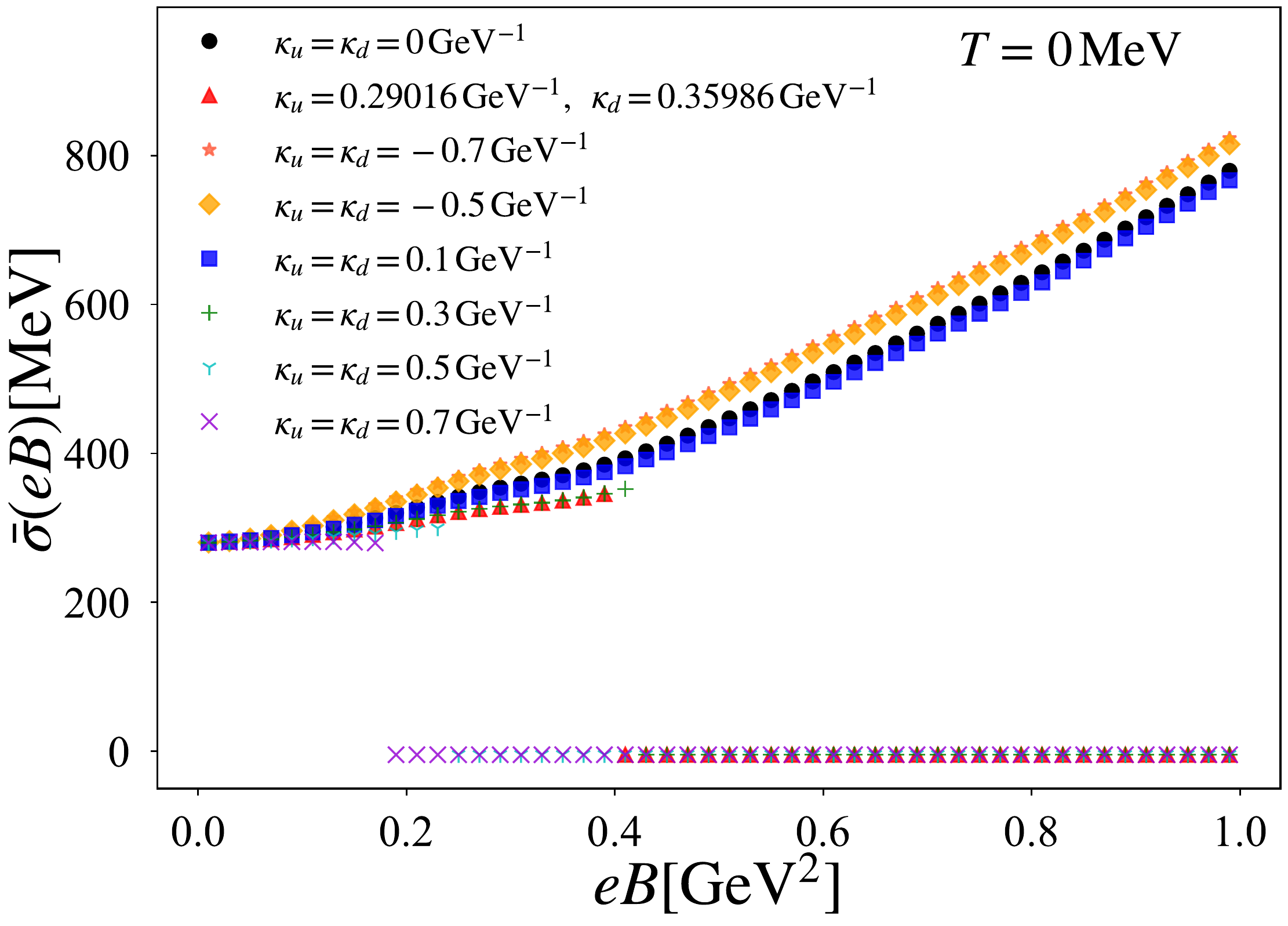}
    \subfigure{(a)}
\end{center}
\end{minipage}
\begin{minipage}{0.5\hsize}
\begin{center}
    \includegraphics[width=8cm]{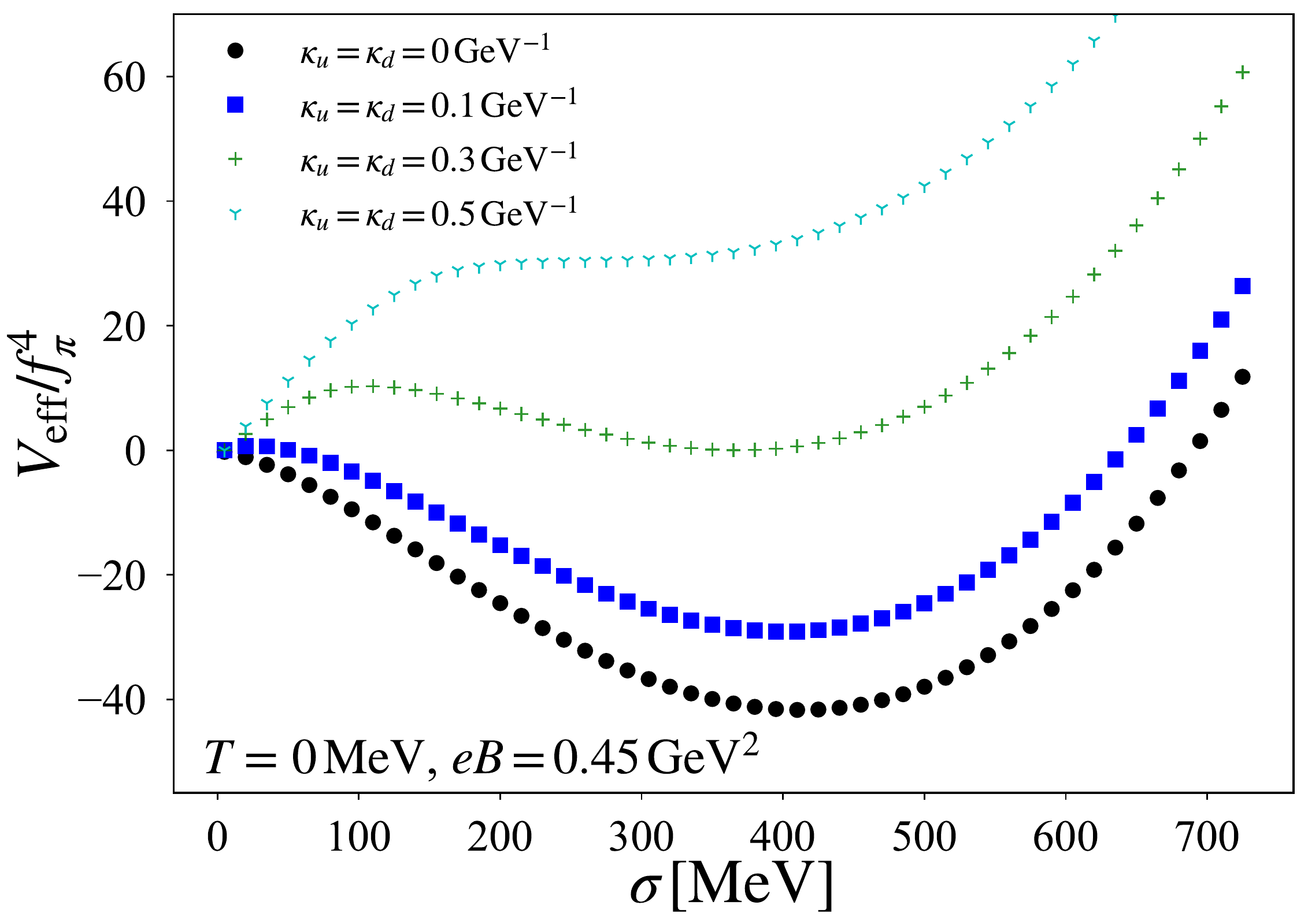}
    \subfigure{(b)}
\end{center}
\end{minipage}
\end{tabular}
\caption{
The constant quark AMM effect on the chiral condensate at $T=0$.
The panel (a) shows the magnetic dependence of the chiral condensate with different values of $\kappa_u=\kappa_d$.
The panel (b) displays the effective potential $V_{\rm eff}$ normalized by the fourth power of the pion decay constant $f_{\pi}$ in the vacuum with $T=0,B=0$. The potentials are also normalized by subtracting $V_{\rm eff}(\sigma=0)$ at $T=0$ MeV. This definition of normalized effective potential follows in the following figures.
}
\label{constAMM_T0}
\end{figure}

In Fig.~\ref{constAMM_T0},
we 
show the constant AMM effect on the the chiral condensate at the zero-temperature.
Note that the magnetic effect on the chiral condensate depends on the regularization procedures for the ultraviolet divergence. Indeed, by using an unsuitable ultraviolet-regulator, the chiral condensate (or the quark condensate) becomes oscillating with the increase of the magnetic field \cite{Avancini:2019wed}.  However, the oscillation in the (subtracted) quark condensate is not observed in the lattice QCD simulations under the constant magnetic field (at $T=0$)~\cite{Bali:2012zg,Bali:2013esa,Ding:2020hxw}. 
To avoid this accidental oscillation in this study, we have chosen the Lorenztian form factor in Eq.~(\ref{regulator}).

In the absence of the AMM,
the chiral condensate $\bar \sigma$ monotonically increases as the magnetic scale gets larger. 
Switching on the constant AMM with the typical values $\kappa_u\simeq 0.29\, {\rm GeV}^{-1}$ and $\kappa_d\simeq 0.36\, {\rm GeV}^{-1}$, a jump arises in the chiral condensate at the critical magnetic field $eB_c\simeq 0.4\,{\rm GeV}^2$. This jump indicates that the NJL model undergoes a first order chiral phase transition induced by the constant AMM. 
The similar behavior of the induced-first order phase transition has been observed in another analysis of the NJL model
\cite {Fayazbakhsh:2014mca}, in which the unphysical oscillation appears in the chiral condensate due to choosing the Woods-Saxon type form factor.  

The constant AMM unexpectedly triggers off the first order phase transition even at the zero temperature system ($T=0$). To clearly understand the induced-first order phase transition, we  make a detailed investigation of the constant AMM effect in the NJL model. For simplicity we suppose that the constant AMM takes 
the isospin symmetric form
$\kappa_u=\kappa_d$, and varied from $-0.7 \,{\rm GeV}^{-1}$ to $0.7 \,{\rm GeV}^{-1}$ in this study. 

The panel (a) of Fig.~\ref{constAMM_T0} shows that, 
for negative constant quark AMM $\kappa_{u,d}<0$,
the monotonic increase of $\bar \sigma$ tends to be enhanced by the constant AMM contribution, and the jump does not show up in the chiral condensate. 
The negative constant quark AMM $\kappa_{u,d}<0$ acts as a catalyzer for the chiral symmetry breaking under a magnetic field, which is similar to the case of tensor-type spin polarization induced by magnetic field as in \cite{Lin:2022ied}. However,  to reconcile the NJL result with the lattice QCD observation,
the catalysis behavior induced by the negative constant quark AMM is not a valuable result. 
For this reason, we discard the negative constant quark AMM  contribution for $\kappa_{u,d}<0$ in the present NJL analysis.

Following that, we consider the case of positive constant quark AMM $\kappa_{u,d}>0$.
For small value, $\kappa_u=\kappa_d=0.1 \,{\rm GeV}^{-1}$, 
the jump does not appear in the chiral condensate, as seen from the panel (a) of Fig.~\ref{constAMM_T0}.
However, 
as the constant AMM $\kappa_{u,d}$ 
becomes larger, the jump suddenly arises. 
In addition, the critical magnetic field corresponding to the jump is shifted to smaller values with the increase of the constant AMM.


Incidentally,
the behavior of the phase transition can be directly seen from the effective potential in Eq.~(\ref{V_eff}).
The panel (b) of Fig.~\ref{constAMM_T0} exhibits a sketch of the constant AMM effect on the effective potential at $T=0$ and $eB=0.45\;{\rm GeV}^2$.
As clearly seen, the potential structure is significantly deformed by the presence of the constant AMM. 
The constant AMM creates the potential barrier between the  chiral symmetric vacuum ($\sigma=0$) and the chiral broken vacuum ($\sigma\neq0$), 
and push up the minimum point at the chiral broken vacuum. 
Due to the significant deformation of the effective  potential, 
the chiral first order phase transition is accidentally induced even at the zero-temperature.  

\begin{figure}[H] 
\begin{tabular}{cc}
\begin{minipage}{0.5\hsize}
\begin{center}
    \includegraphics[width=8cm]{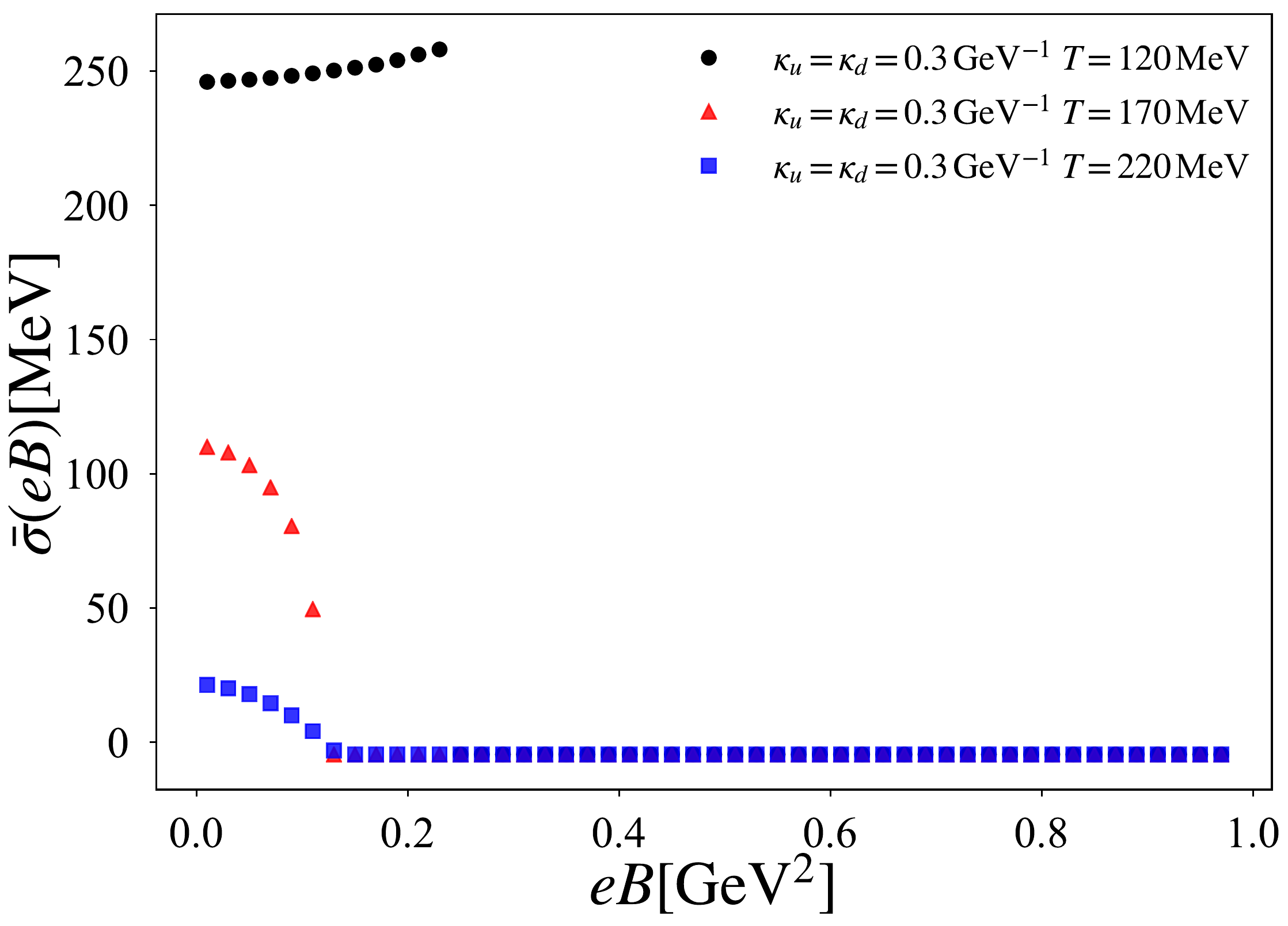}
    \subfigure{(a)}
\end{center}
\end{minipage}
\begin{minipage}{0.5\hsize}
\begin{center}
    \includegraphics[width=8cm]{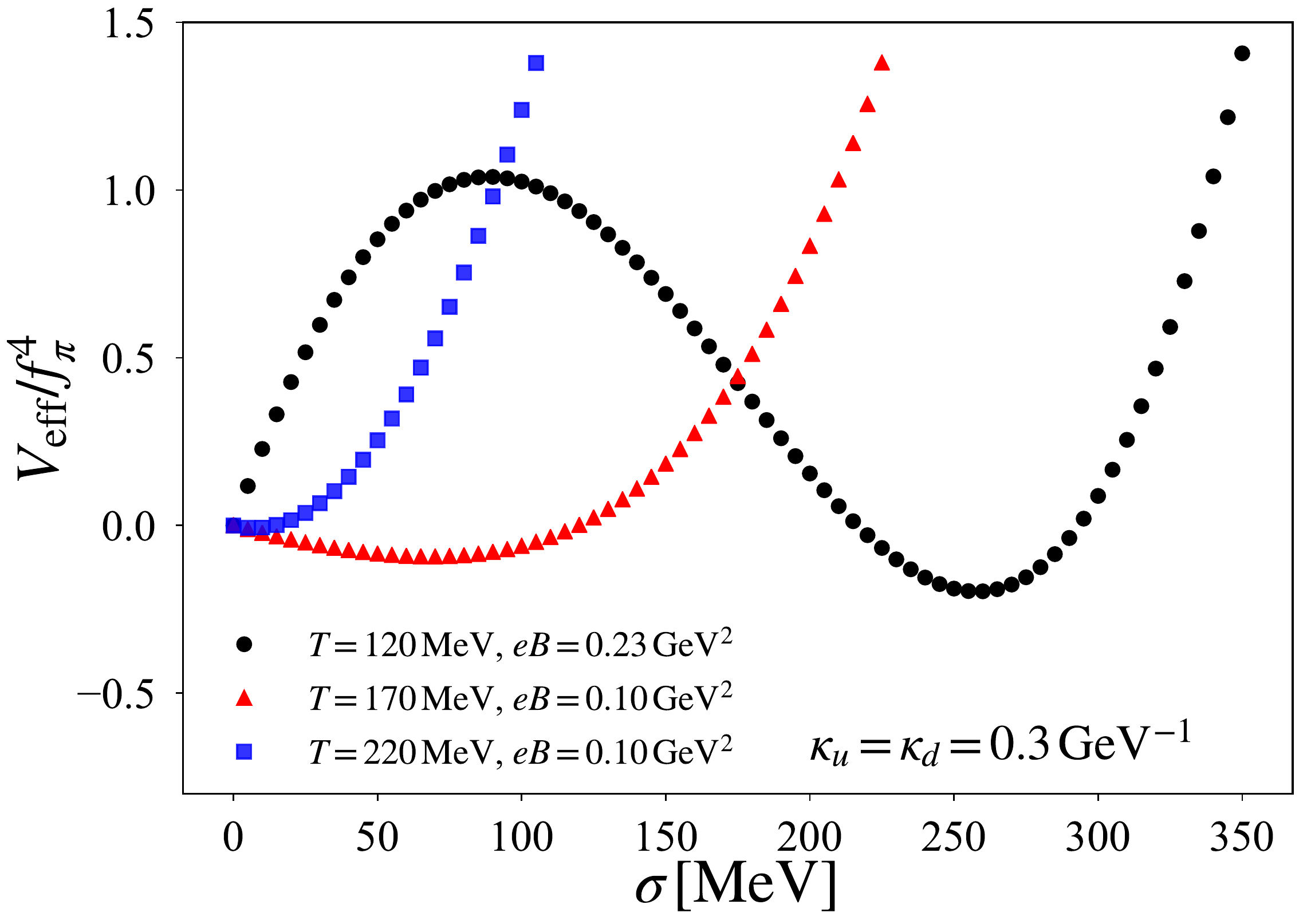}
    \subfigure{(b)}
\end{center}
\end{minipage}
\end{tabular}
\caption{
The chiral phase transition under external magnetic fields with a constant quark AMM in the case of $\kappa_u=\kappa_d=0.3\,{\rm GeV}^{-1}$ for different temperatures. The panel (a): the chiral phase transition along the magnetic field with different temperatures. The panel (b): the deformation of the normalized effective potential by the pion decay constant $f_{\pi}$ in the vacuum. 
}
\label{constAMM_finiteT_vev_pot}
\end{figure}

Next we move onto the thermal system. 
Figure~\ref{constAMM_finiteT_vev_pot}
shows the thermal effect on the chiral phase transition along the magnetic field, which is induced by the constant AMM. Here we take the form of $\kappa_u=\kappa_d=0.3\,{\rm GeV}^{-1}$ as one of the cases for the induced-first order phase transition at the zero-temperature.
At low temperatures below the the critical temperature $T<T_{\rm pc}^{(\rm NJL)}(eB=0)\, (\simeq 167\,{\rm MeV})$,
the chiral phase transition still keeps of
 first order due to the presence of the potential barrier, as depicted in the panel (b) of Fig.~\ref{constAMM_finiteT_vev_pot}. However, the potential barrier is washed out by the thermal effect at high temperatures above the critical temperature $T>T_{\rm pc}^{\rm (NJL)}(eB=0)$. As a result, the chiral phase transition changes to be of a second order phase transition. 

\begin{figure}[H] 
\begin{tabular}{cc}
\begin{minipage}{0.33\hsize}
\begin{center}
    \includegraphics[width=5.8cm]{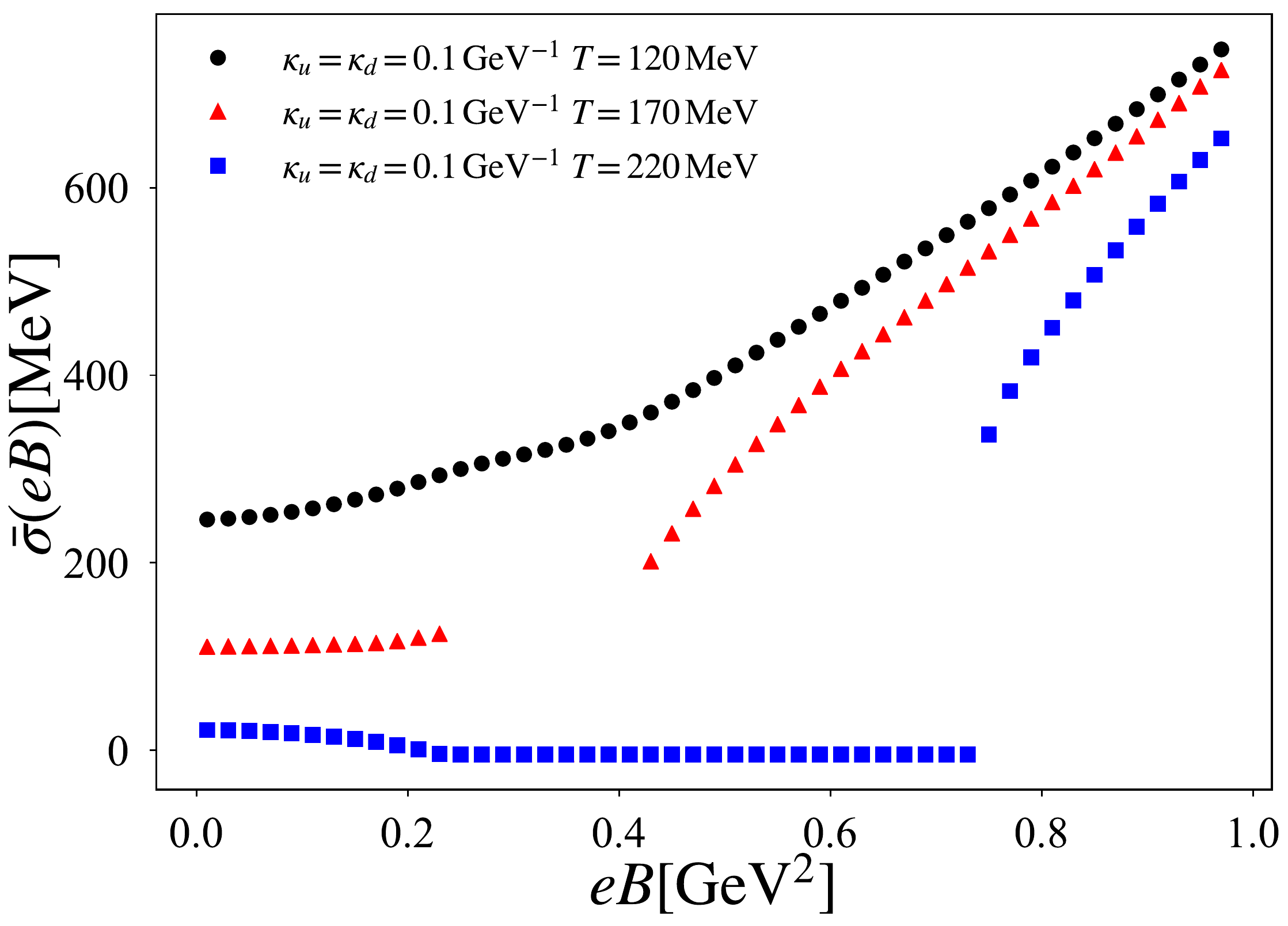}
    \subfigure{(a)}
\end{center}
\end{minipage}
\begin{minipage}{0.33\hsize}
\begin{center}
    \includegraphics[width=5.8cm]{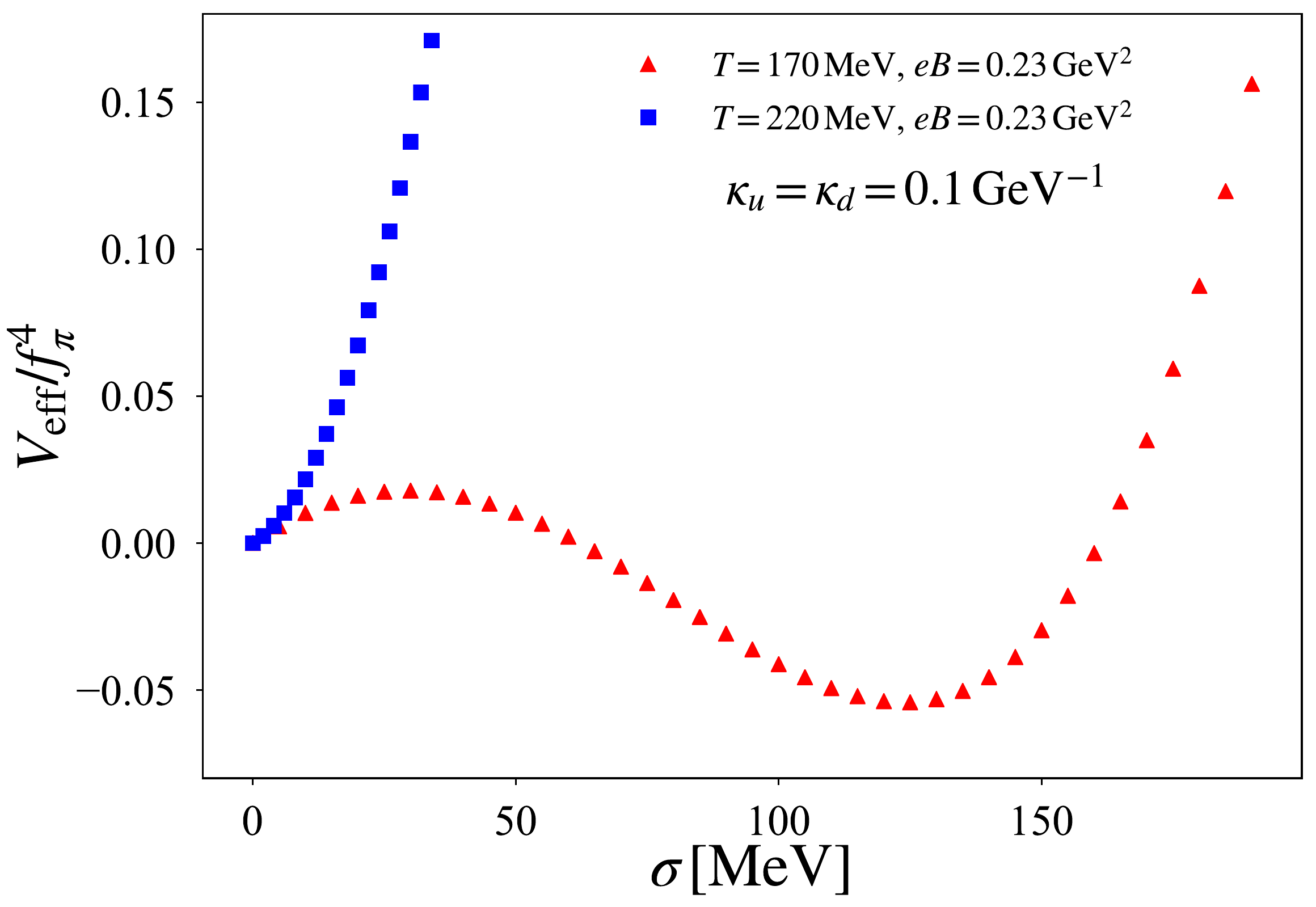}
    \subfigure{(b)}
\end{center}
\end{minipage}
\begin{minipage}{0.33\hsize}
\begin{center}
    \includegraphics[width=5.8cm]{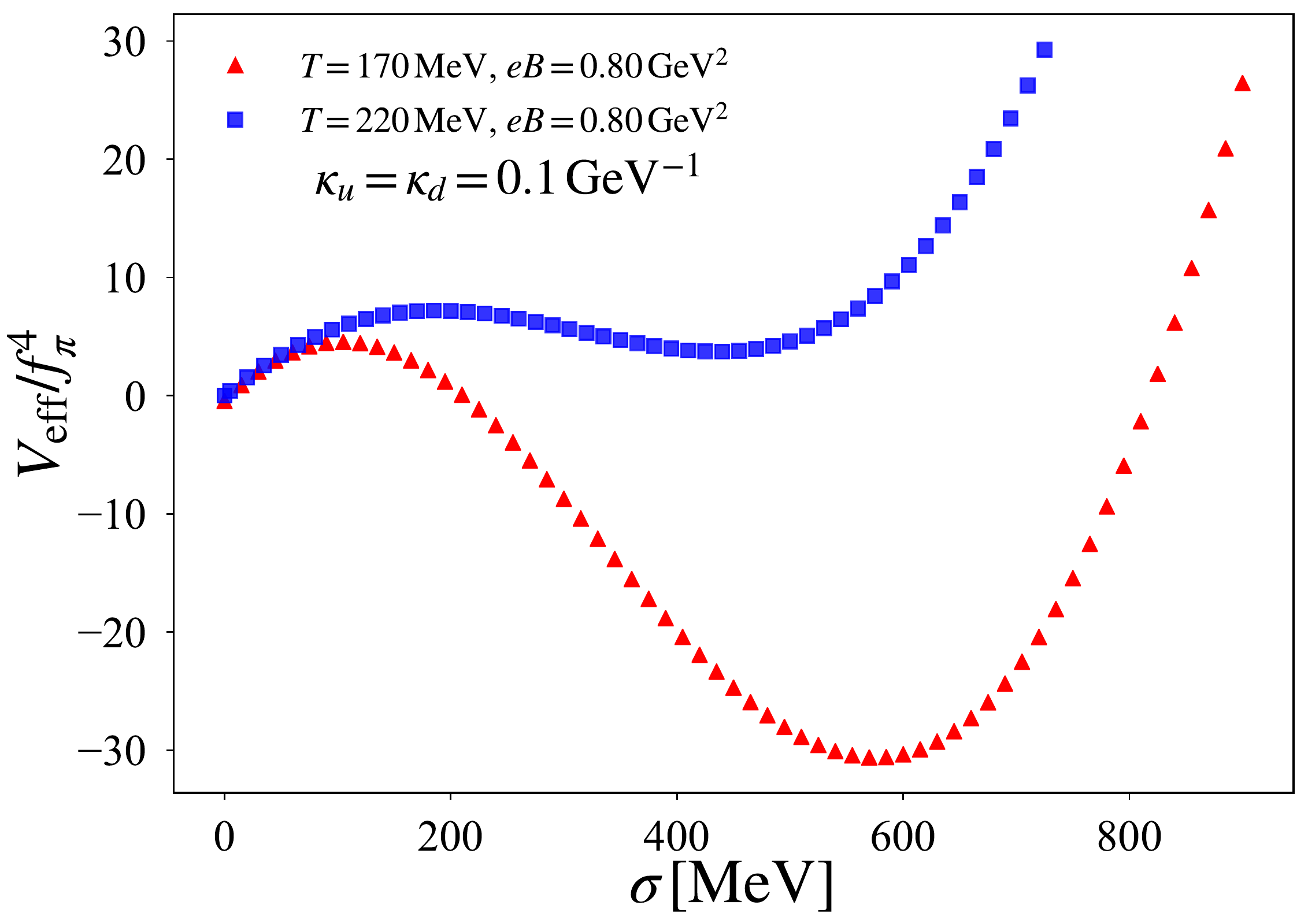}
    \subfigure{(c)}
\end{center}
\end{minipage}
\end{tabular}
\caption{
Similar to Fig.~\ref{constAMM_finiteT_vev_pot}, but in the case of $\kappa_u=\kappa_d=0.1\,{\rm GeV}^{-1}$. The normalized effective potential at $eB=0.23\,{\rm GeV}^{2}$ in (b) and $eB=0.8\,{\rm GeV}^{2}$ in (c).
}
\label{constAMM_finiteT}
\end{figure}

As mentioned above in the case of the zero-temperature, the small-constant AMM corresponding to $\kappa_u=\kappa_d=0.1 \,{\rm GeV}^{-1}$ hardly contributes to the chrial condensate. However, around the critical temperature $T=T_{\rm pc}^{\rm (NJL)}(eB=0)\, (\simeq 167\,{\rm MeV})$, the small-constant AMM becomes eminent, and the order of the chiral phase transition drastically changes. Fig.~\ref{constAMM_finiteT} shows that even with a small-constant AMM
a jump shows up and a first order phase transition is realized 
at around $T=T_{\rm pc}^{(\rm NJL)}(eB=0)$. Particularly, it is interesting to note that there exist two jumps at around $eB_c\simeq 0.23\,{\rm GeV}^{2}$ and $eB_c\simeq 0.4\,{\rm GeV}^{2}$ for $T=170\,{\rm MeV}$, which might be due to the competition between the magnetic catalysis and the magnetic inhibition. 
In fact, the potential barrier is formed at around the critical magnetic fields, as seen from the panel (b) and (c) of Fig.~\ref{constAMM_finiteT}. 
As the temperature increases, the potential barrier corresponding to $eB_c\simeq 0.23\,{\rm GeV}^{2}$ vanishes, so that  
the phase transition changes to the second order phase transition. On the other hand, the potential barrier corresponding to $eB_c\simeq 0.4\,{\rm GeV}^{2}$ remains even at higher temperature ($T=220$ MeV), but the critical magnetic field is shifted to stronger magnetic field region, as shown in panel (a) in Fig.~\ref{constAMM_finiteT}.



 Note that the chiral phase transitions including the constant AMM effect at the finite temperature have already been discussed in the NJL model \cite{Fayazbakhsh:2014mca,Chaudhuri:2019lbw} and the PNJL model~\cite{Mei:2020jzn}. 
However, the analysis on the potential deformation has not explicitly been shown so far.

Due to the presence of the constant AMM,
the phase structure becomes rich. 
However, the result from lattice QCD simulations at the thermomagnetic system yields the chiral crossover \cite{Bali:2011qj,Bali:2012zg,Bali:2013esa,Ding:2020hxw, Ding:2022tqn},
which is inconsistent with this estimate in the NJL model with constant AMM.   
This implies that the constant AMM would not be a suitable form for the low-energy effective theory based on the underlying QCD. 
In the next subsection, we deal with another description of the quark AMM.  

\subsection{Magnetic-dependent AMM-I:
$\kappa_{u,d}=v \sigma$}

\begin{figure}[H]
\begin{tabular}{cc}
\begin{minipage}{0.5\hsize}
\begin{center}
    \includegraphics[width=8cm]{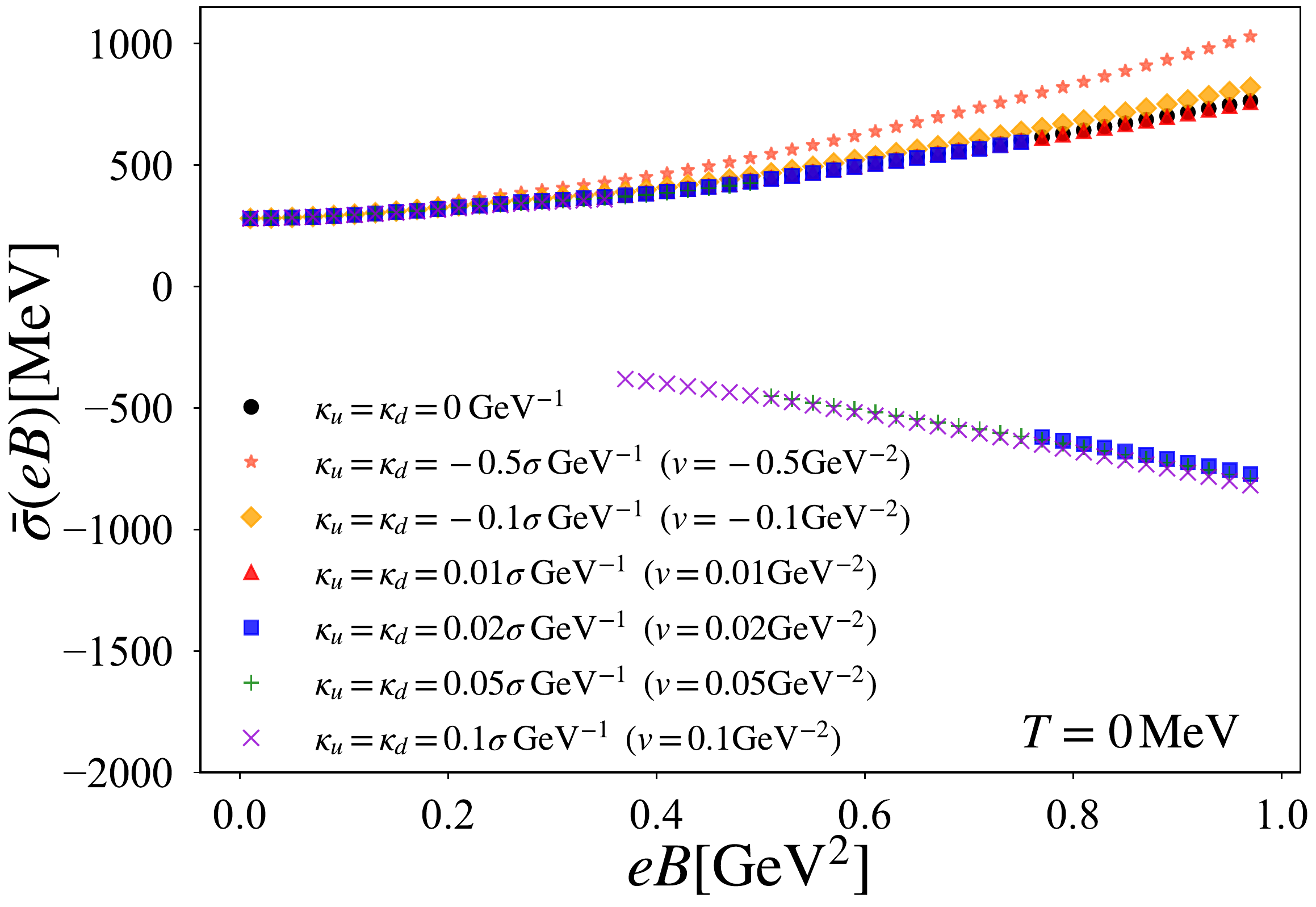}
    \subfigure{(a)}
\end{center}
\end{minipage}
\begin{minipage}{0.5\hsize}
\begin{center}
    \includegraphics[width=8cm]{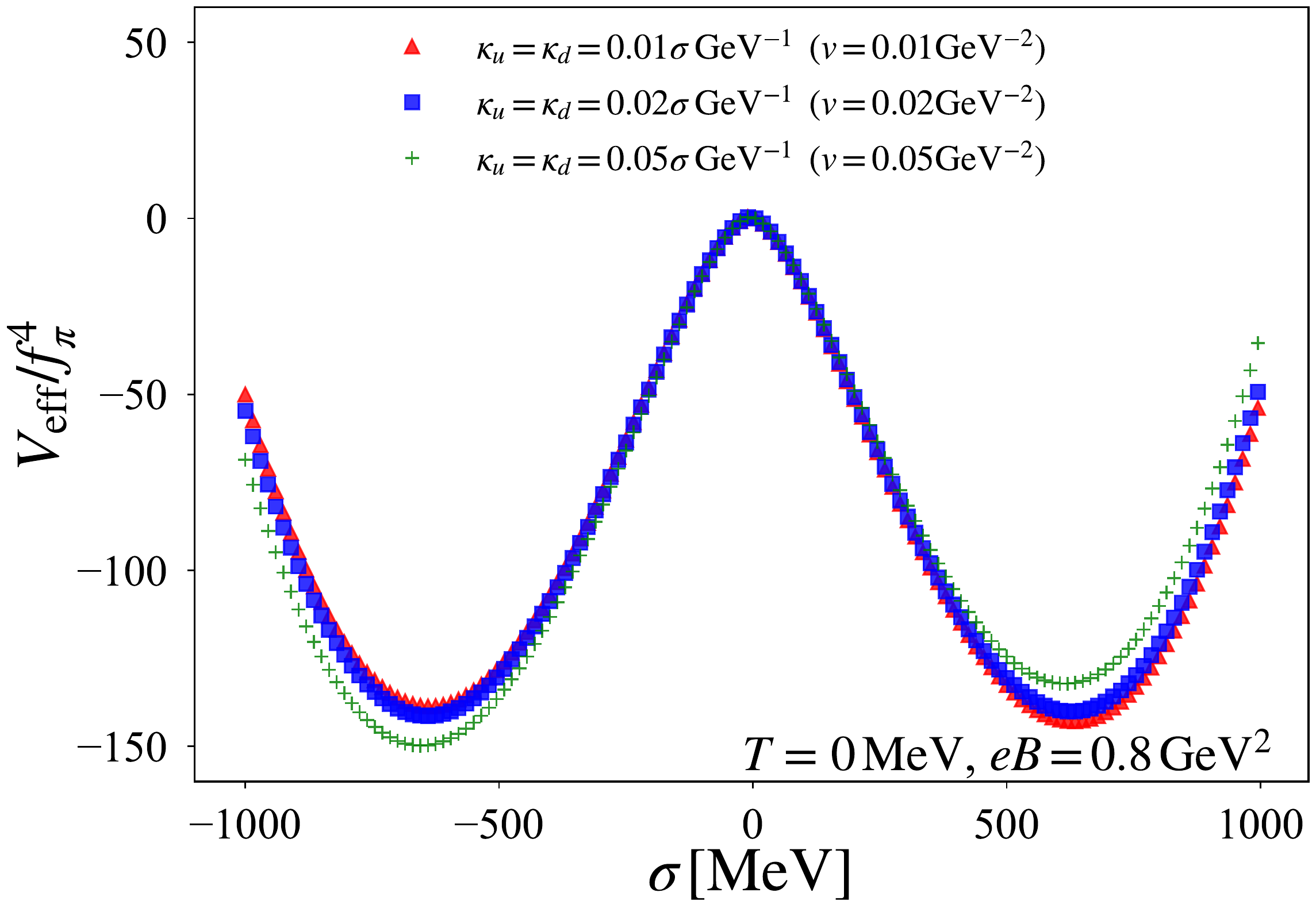}
    \subfigure{(b)}
\end{center}
\end{minipage}
\end{tabular}
\caption{
Similar to Fig.~\ref{constAMM_T0}, but in the case of the magnetic-dependent AMM proportional to the chiral condensate: $\kappa_{u}=\kappa_{d}= v \sigma$.
The panel (a) shows that 
the magnetic-dependent AMM give a flip of the sign on the chiral condensate. 
In the panel (b),
the normalized effective potentials is evaluated at $T=0$ and $eB=0.8~{\rm GeV}^{2}$. This displays that the global minimum point of the normalized effective potential jumps from the positive vacuum to the negative vacuum.
}
\label{TMAMM_case1_T0}
\end{figure}

In this subsection, we consider the magnetic-dependent quark AMM proportional to the chiral condensate: $\kappa_{u}=\kappa_{d}= v \sigma$. Fig.\ref{TMAMM_case1_T0} displays the magnetic-dependent AMM $\kappa_{u}=\kappa_{d}= v \sigma$ effect on the chiral condensate at zero temperature.  
To begin with, we set the AMM parameter $v$ to negative values. 
The panel (a) of Fig.~\ref{TMAMM_case1_T0} shows that,
for $v<0$, 
the magnetic-dependent AMM plays a role of the catalyzer for the chiral symmetry breaking under the magnetic field.
Thus, we also discard the magnetic-dependent AMM contribution for $v<0$ in this study, like in the case of the negative value of the constant AMM parameter ($\kappa_{u,d}={\rm const.}$). Below, we only focus on positive values of the AMM parameter $v$. 

For $v=0.01\,{\rm GeV}^{-2}$, the magnetic scaling of the chiral condensate is almost on the same trajectory as the one without quark AMM, as shown in the panel (a) of Fig.~\ref{TMAMM_case1_T0}. 
As the AMM parameter $v$ grows, for $v=0.02\,{\rm GeV}^{-2}$, the striking phenomena emerge in the chiral condensate, i.e., a jump of $\bar \sigma$ shows up at a critical magnetic field, moreover, the sign of the chiral condensate flips from positive to negative. In addition, the critical magnetic field of $\bar \sigma$ becoming negative shifts to a weaker magnetic field when the AMM parameter $v$ increases. 
This flip of the sign of $\bar \sigma$ can be explicitly viewed from the deformation of the effective potential, as exhibited in the panel (b) of Fig.~\ref{TMAMM_case1_T0}, where the magnetic field is fixed as $eB=0.8~{\rm GeV}^{2}$:
the global minimum point of the effective potential jumps from the positive vacuum ($\bar \sigma>0$) to the negative vacuum ($\bar \sigma<0$) 
as the AMM parameter $v$ increases. 

Note that 
the magnetic-dependent AMM for $\kappa_u=\kappa_d=v\sigma$ has already been addressed for the NJL model analysis 
in \cite{Xu:2020yag,Lin:2022ied}. 
However, the flipped-$\bar \sigma$ or negative $\bar \sigma$ solution has been simply dropped  in \cite{Xu:2020yag,Lin:2022ied}, because a negative $\bar \sigma$ indicates a negative dynamical quark mass, which was believed as unphysical.


\begin{figure}[H]
\begin{tabular}{cc}
\begin{minipage}{0.5\hsize}
\begin{center}
    \includegraphics[width=8cm]{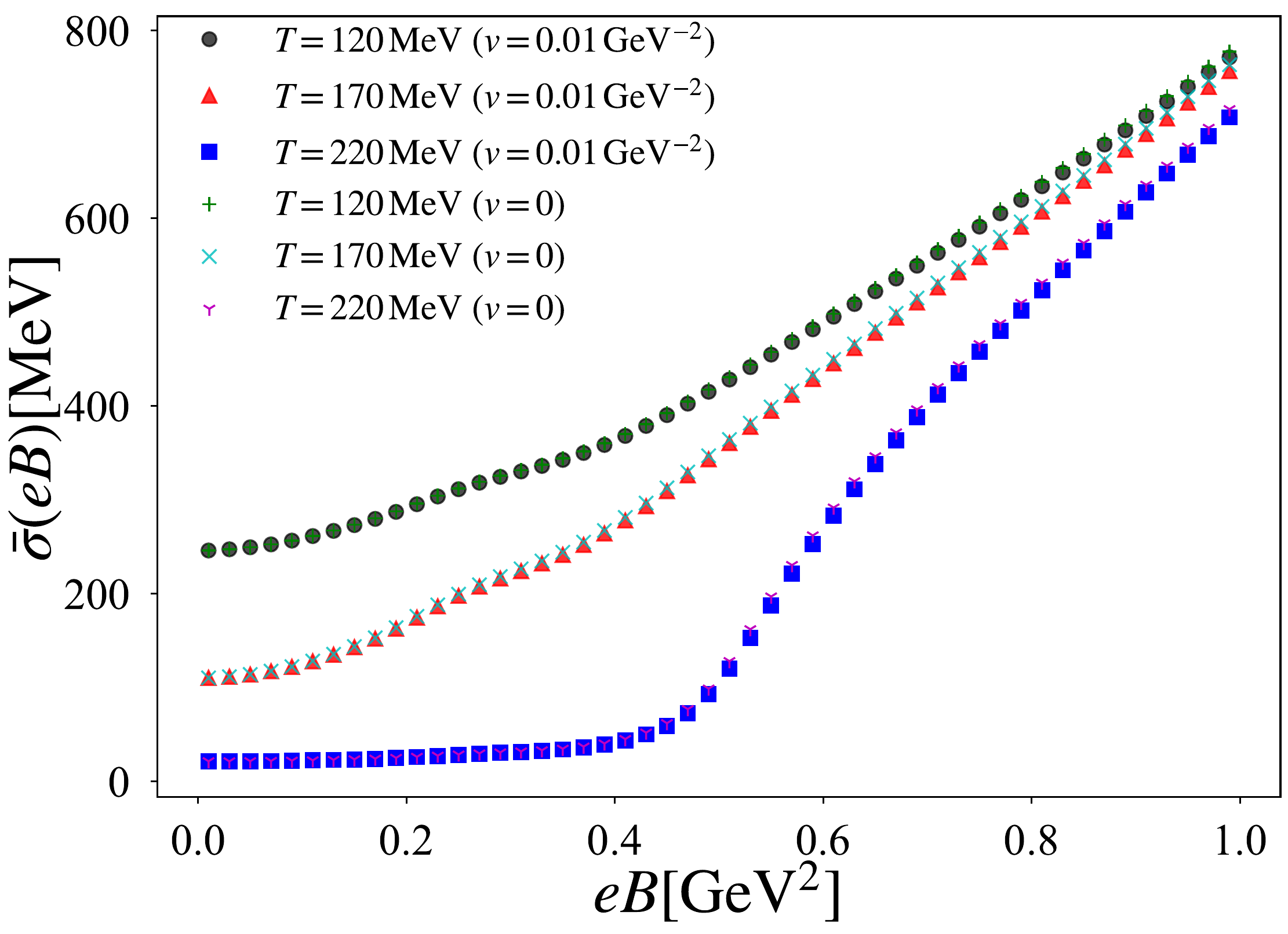}
    \subfigure{(a)}
\end{center}
\end{minipage}
\begin{minipage}{0.5\hsize}
\begin{center}
    \includegraphics[width=8cm]{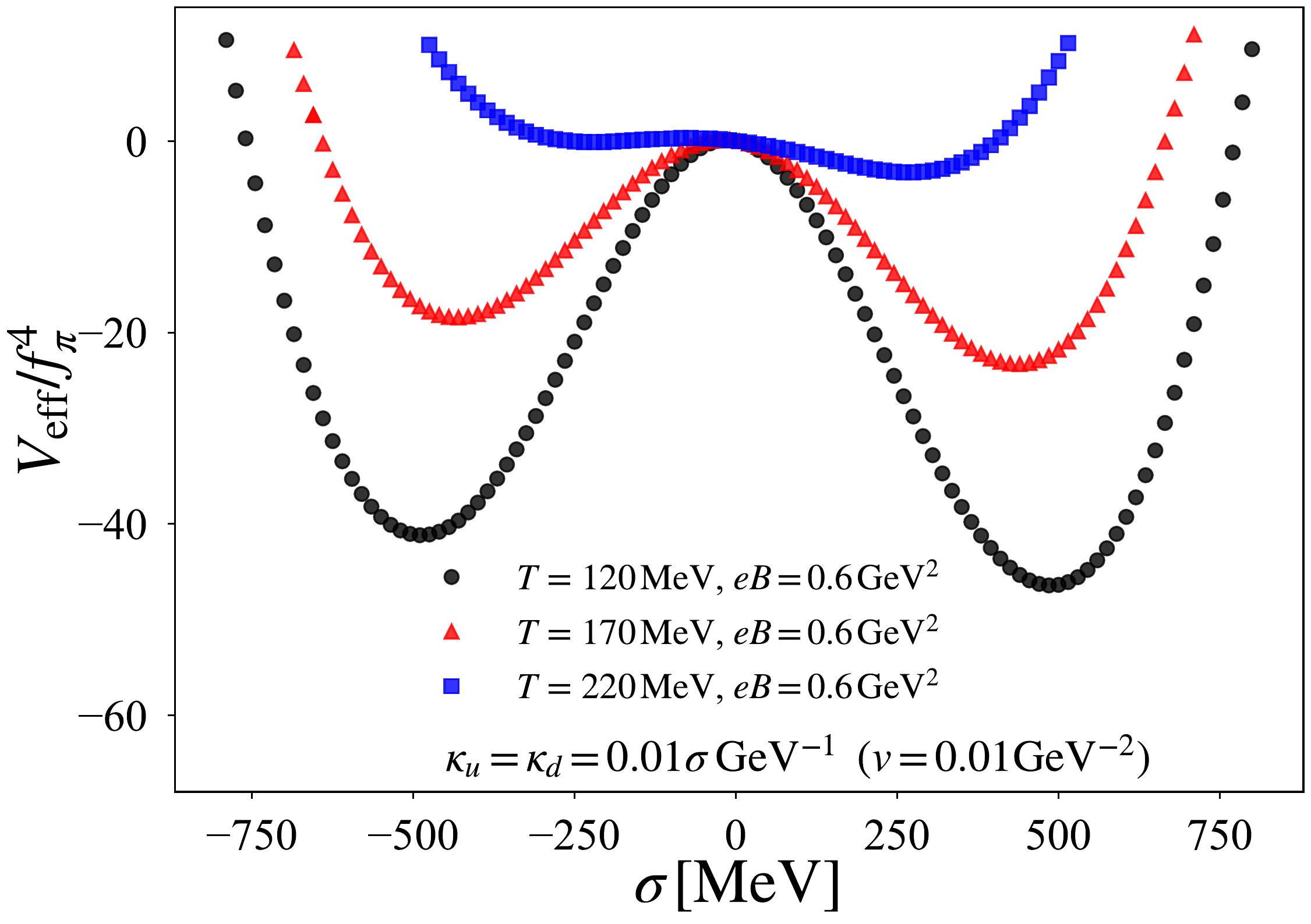}
    \subfigure{(b)}
\end{center}
\end{minipage}
\end{tabular}
\caption{(a):
The thermal effect on the chiral condensate as a function of the magnetic field for $\kappa_{u}=\kappa_d=0.01 \sigma\,{\rm GeV}^{-1}$ $(v=0.01\,{\rm GeV}^{-2})$, which is compared with the case without the quark AMM contribution.
(b): The corresponding normalized effective potentials for $v=0.01\,{\rm GeV}^{-2}$ are evaluated at $eB=0.6\,{\rm GeV}^{2}$.
}
\label{TMAMM_case1_T_k001}
\end{figure}

We also evaluate the thermal effect on the chiral condensate including the magnetic-dependent AMM effect. 
From the panel (a) of Fig.~\ref{TMAMM_case1_T_k001}, we find that, in the case of the small AMM parameter corresponding to $v=0.01 \,{\rm GeV}^{-2}$,  
the magnetic-dependent AMM  hardly contributes to the chiral condensate even at finite temperatures. 
The monotonic increasing $\bar \sigma$ for $v=0.01 \,{\rm GeV}^{-2}$ persists and the flip on the sign of $\bar \sigma$ dose not appear for any temperatures. 

\begin{figure}[H]
\begin{tabular}{cc}
\begin{minipage}{0.5\hsize}
\begin{center}
    \includegraphics[width=8cm]{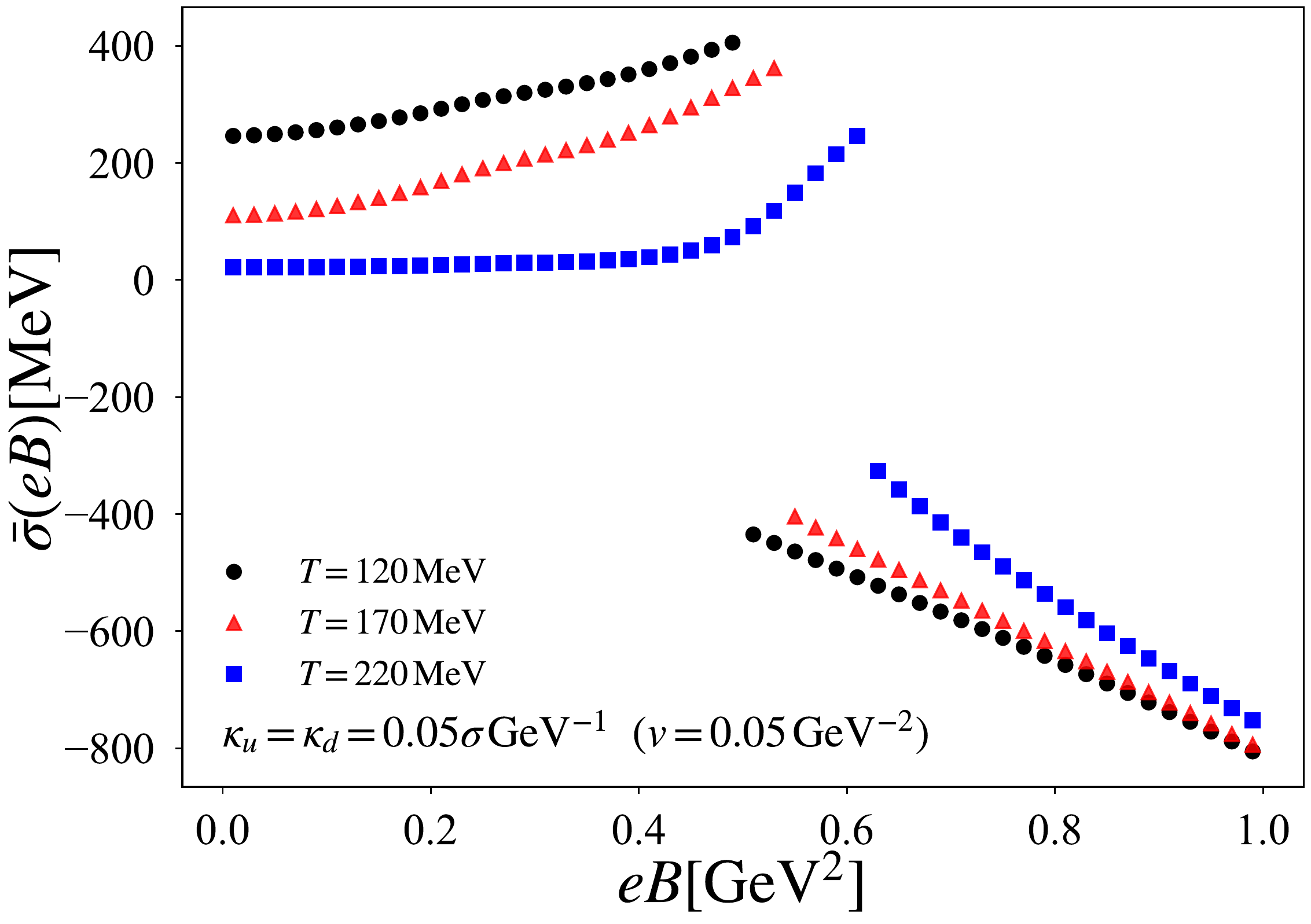}
    \subfigure{(a)}
\end{center}
\end{minipage}
\begin{minipage}{0.5\hsize}
\begin{center}
    \includegraphics[width=8cm]{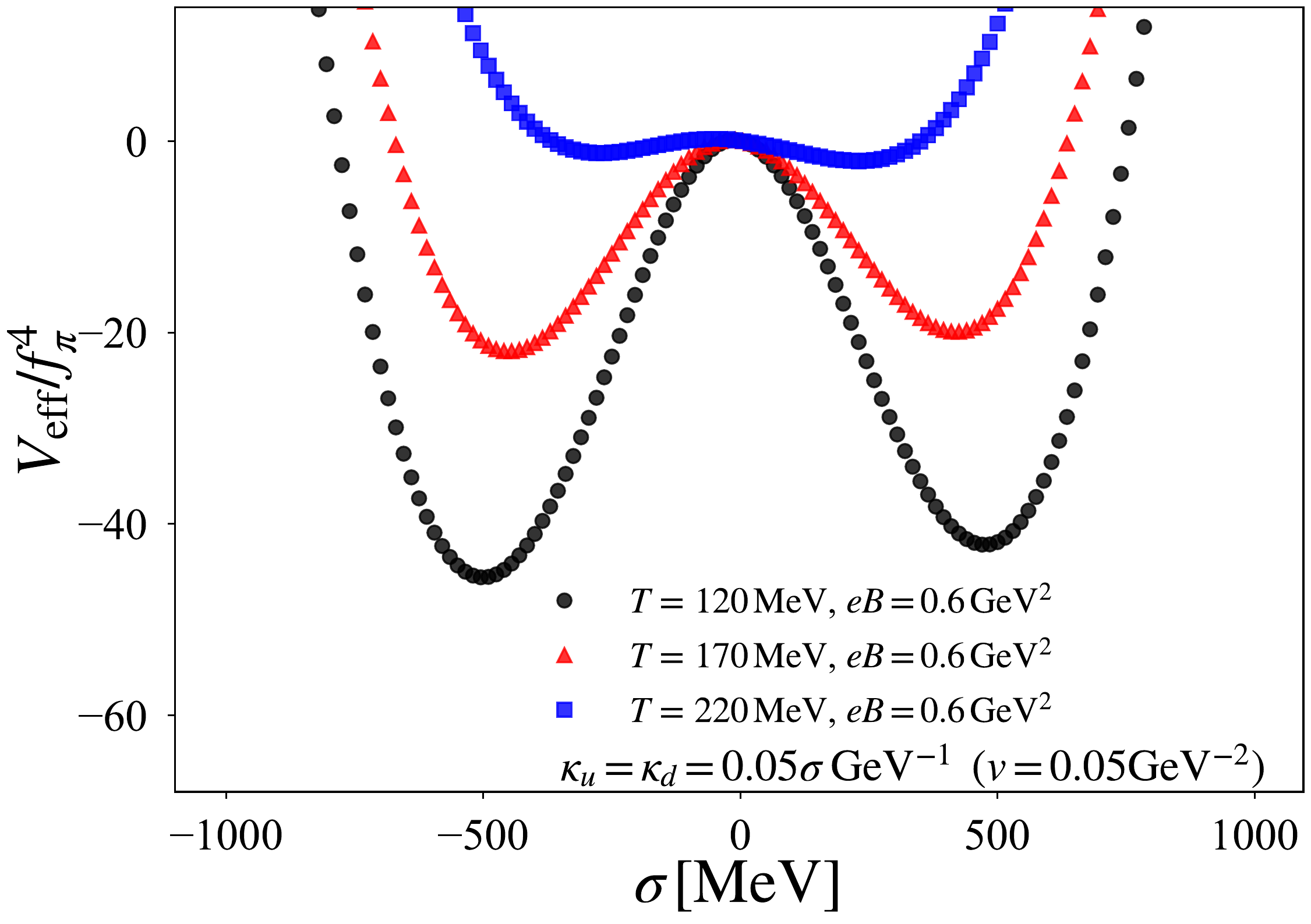}
    \subfigure{(b)}
\end{center}
\end{minipage}
\end{tabular}
\caption{
(a): The thermal effect on the chiral condensate as a function of the magnetic field for $\kappa_{u}=\kappa_d=0.05 \sigma\,{\rm GeV}^{-1}$ $(v=0.05\,{\rm GeV}^{-2})$.
(b): the corresponding normalized effective potentials are evaluated at $eB=0.6\,{\rm GeV}^{-2}$.
}
\label{TMAMM_case1_T_k005}
\end{figure}

For $v=0.05\,{\rm GeV}^{-2}$ as a case of the flipped-$\bar \sigma$, the finite temperature interferes with the flip on $\bar \sigma$.   
The flip-point is shifted to a stronger magnetic field region with the increase of the temperature  
(see the panel (a) of Fig.~\ref{TMAMM_case1_T_k005}).
This behavior can also be seen from the deformation of the effective potential.
In the panel (b) of Fig.~\ref{TMAMM_case1_T_k005} where the magnetic field is fixed as $eB=0.6\,{\rm GeV}^{2}$, 
the effective potential structure is deformed by the thermal effect, so that the global minimum point of the effective potential is put back on the positive vacuum at high temperatures.

Although the flip on the sign of the chiral condensate may possibly induce intriguing phenomena in the meson dynamics,
the flipped-chiral condensate (quark condensate) 
has not been seen in the lattice QCD simulations~\cite{Bali:2011qj,Bali:2012zg,Bali:2013esa,Ding:2020hxw, Ding:2022tqn}.
Thus, the magnetic dependent AMM proportional to the chiral condensate would be also an unsuitable effective form to describe the QCD vacuum under the constant magnetic field.

\subsection{
Magnetic-dependent AMM-II: $\kappa_{u,d}=\bar v \sigma^2$}\label{amm2}
\begin{figure}[H] 
\begin{tabular}{cc}
\begin{minipage}{0.5\hsize}
\begin{center}
    \includegraphics[width=8cm]{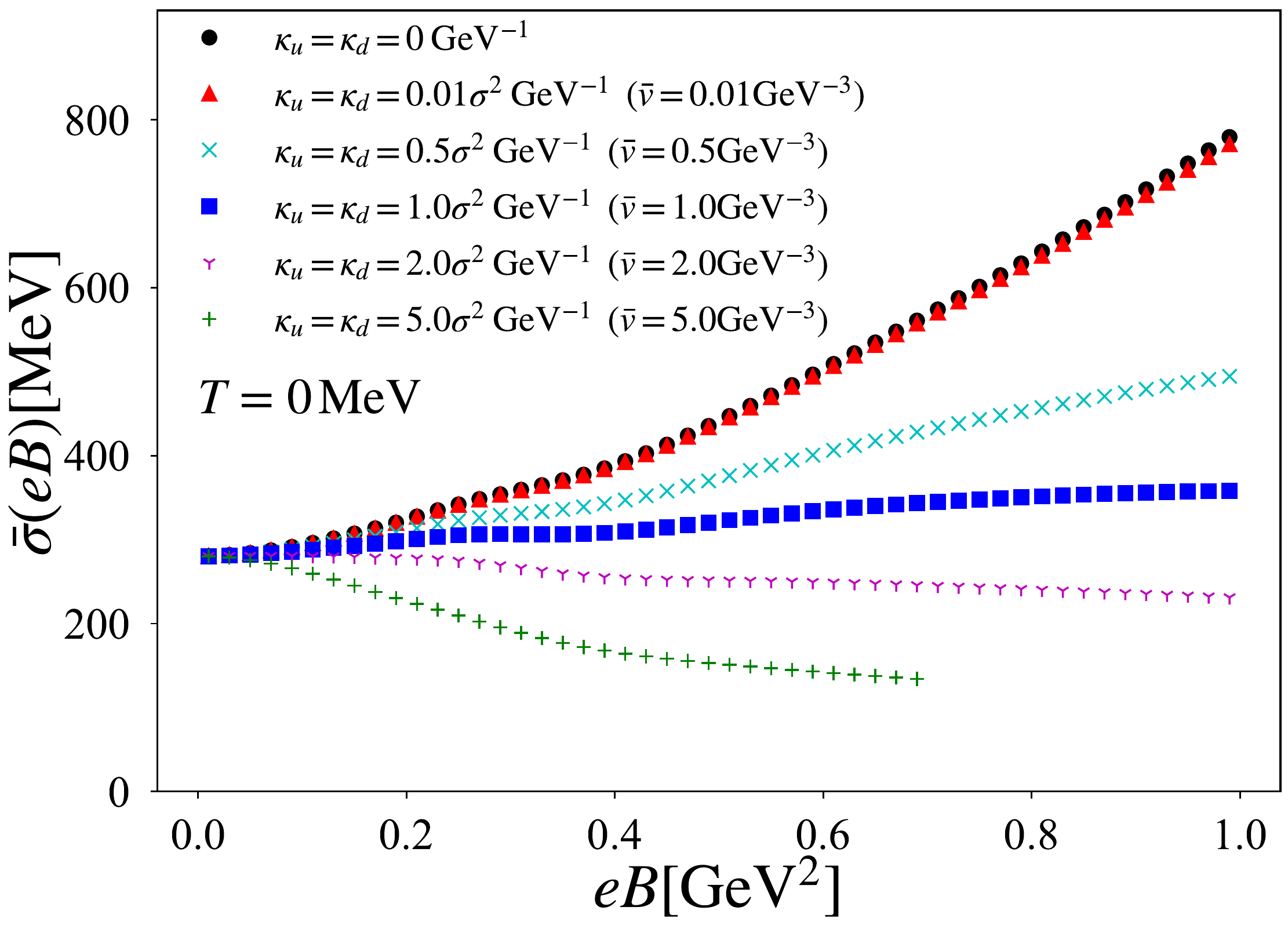}
    \subfigure{(a)}
\end{center}
\end{minipage}
\begin{minipage}{0.5\hsize}
\begin{center}
    \includegraphics[width=8cm]{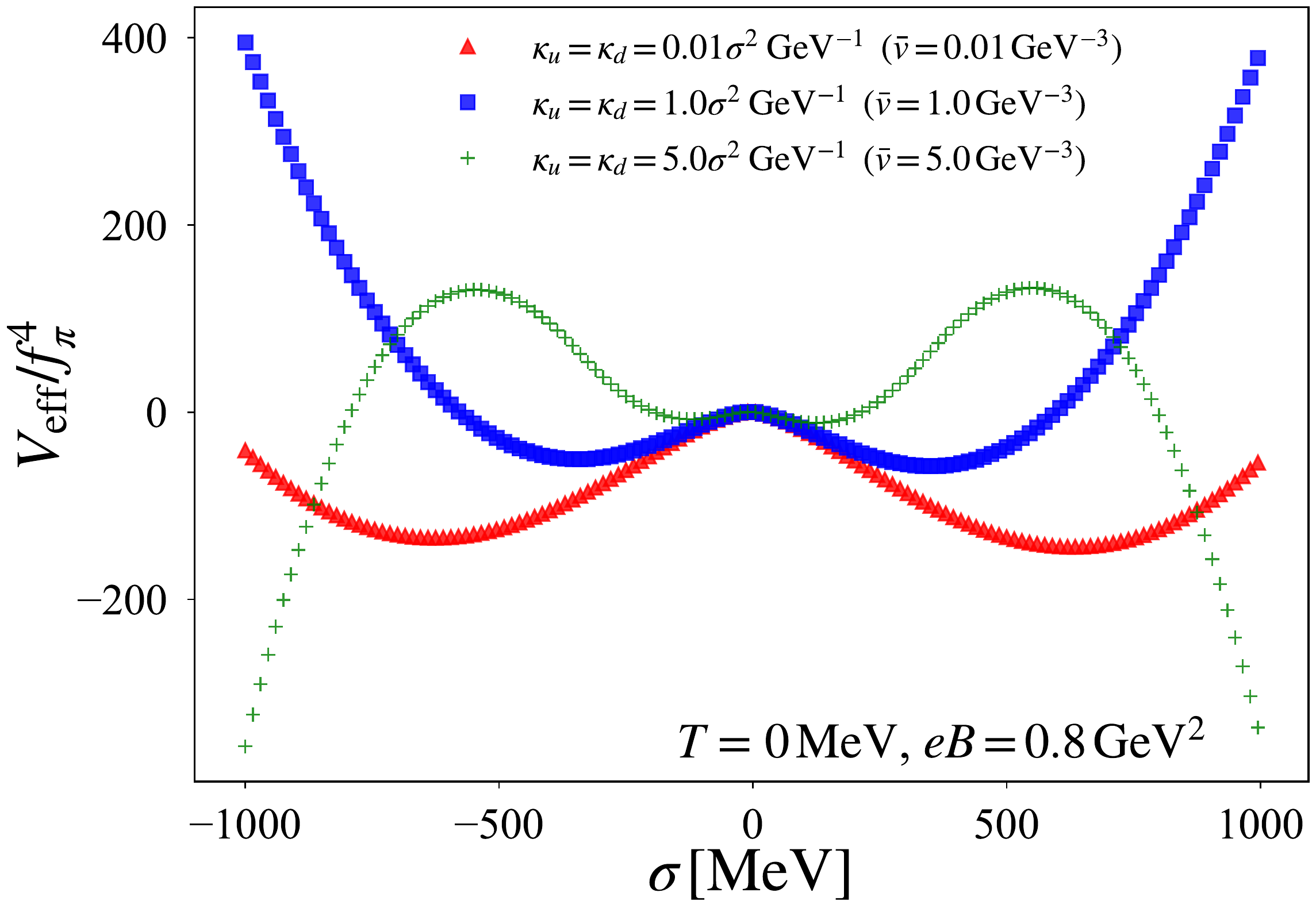}
    \subfigure{(b)}
\end{center}
\end{minipage}
\end{tabular}
\caption{
Similar to Fig.~\ref{TMAMM_case1_T0}
but in the case of the magnetic-dependent AMM proportional to the square of the chiral condensate: $\kappa_{u}=\kappa_{d}= \bar v \sigma^2$. 
The panel (a): the thermal effect on the chiral condensate as a function of the magnetic field. 
The panel (b): the normalized effective potential evaluated at $T=0$ and $eB=0.8~{\rm GeV}^{2}$.
}
\label{MAMM2_T0}
\end{figure}

Now
we take the magnetic-dependent AMM for $\kappa_u=\kappa_d =\bar v\sigma^2$
\footnote{
For $\bar v<0$, 
the magnetic-dependent AMM ($\kappa_{u,d}=\bar v \sigma^2$) promotes the enhancement of the chiral symmetry breaking under the magnetic field.
For the same reason in $\kappa_{u,d}={\rm const.}$ and $\kappa_{u,d}=v\sigma$, we omit the negative value for the AMM parameter $\bar v$ in the present study.
}. Figure~\ref{MAMM2_T0} illustrates a similar plot as Fig.~\ref{TMAMM_case1_T0}.
This figure shows that, at zero temperature, the development of the AMM parameter $\bar v$ suppresses 
the enhancement of the chiral condensate with the increase of the magnetic field.
Therefore
the magnetic-dependent AMM
for $\bar v>0$ 
plays the role of the destructive interference or magnetic inhibition for the chiral symmetry breaking under the magnetic field.

However, when the AMM parameter $\bar v$ takes a value larger than $2~{\rm GeV}^{-3}$, 
the vacuum structure of the NJL model becomes unstable in the sufficient magnetic field region.
As sketched in the panel (b) of Fig.~\ref{MAMM2_T0}, 
the magnetic-dependent AMM for $\bar v=5.0~{\rm GeV}^{-3}$
catastrophically deforms the effective potential at a strong magnetic field $eB=0.8{\rm GeV}^2$. 
Therefore the vacuum structure collapses due to the significant contribution of the  magnetic-dependent AMM, and then
the global minimum of the effective potential goes away.  
 Consequently, in the panel (a) of Fig.~(\ref{MAMM2_T0}),
 the chiral condensate in the case of $\bar v=5.0~{\rm GeV}^{-3}$ vanishes 
for $eB\geq0.7{\rm GeV}^2$.

\begin{figure}[H] 
\begin{tabular}{cc}
\begin{minipage}{0.5\hsize}
\begin{center}
    \includegraphics[width=8cm]{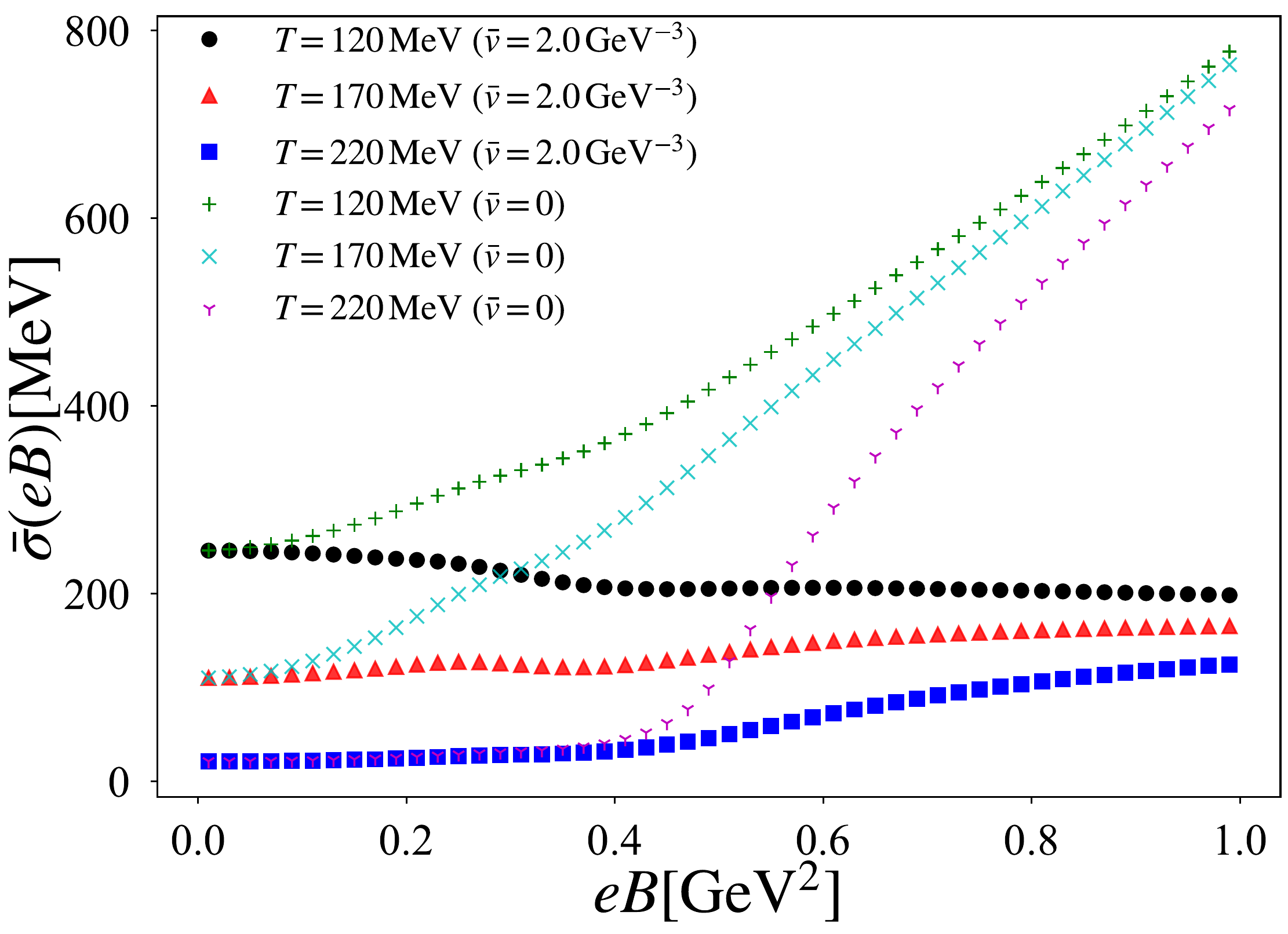}
    \subfigure{(a)}
\end{center}
\end{minipage}
\begin{minipage}{0.5\hsize}
\begin{center}
\includegraphics[width=8cm]{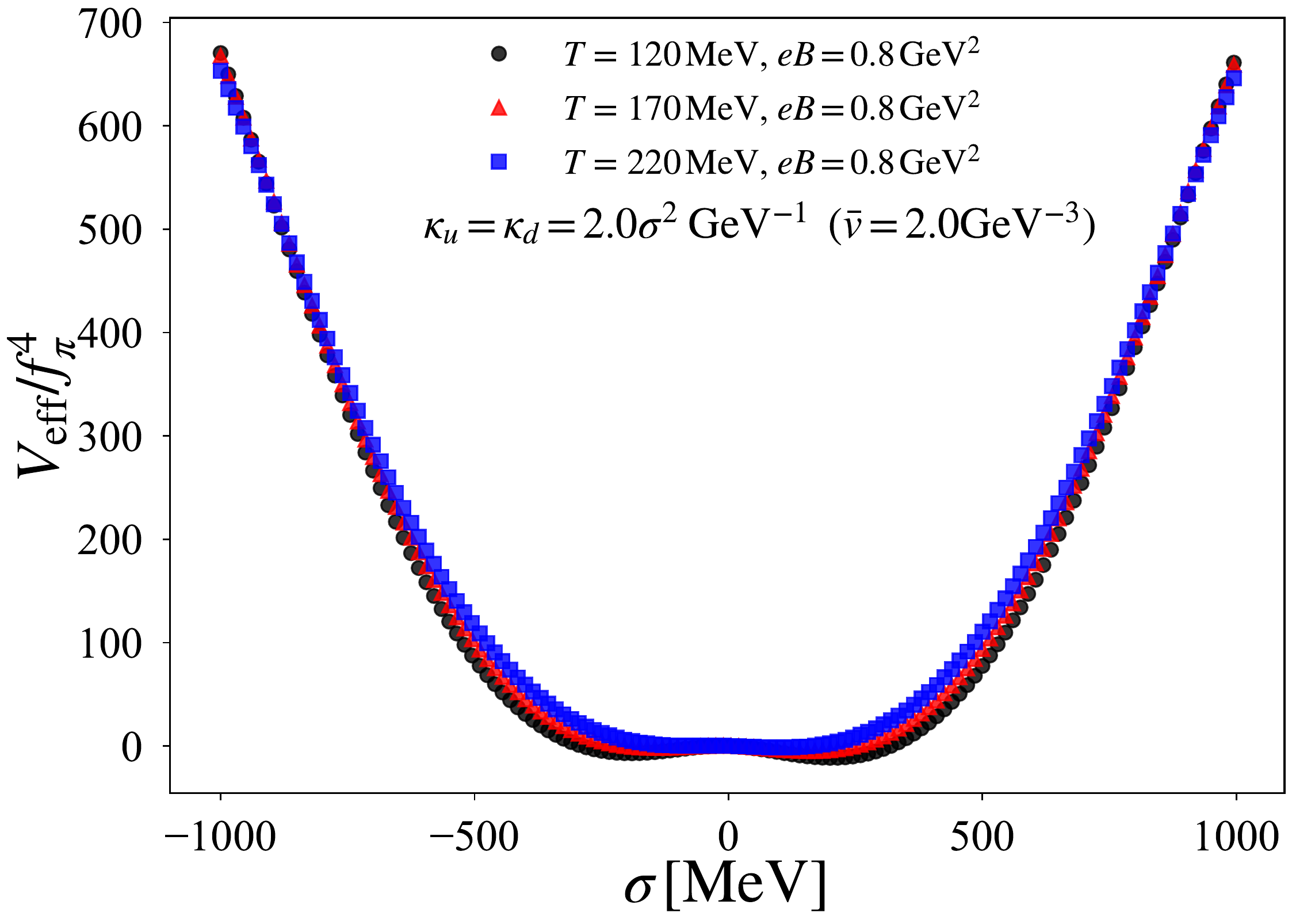}
    \subfigure{(b)}
\end{center}
\end{minipage}
\end{tabular}
\caption{
(a): The thermal effect on the chiral condensate for $\bar v=2.0{\rm GeV}^{-3}$, in comparison with the case without the quark AMM.
(b): The thermal deformation of the normalized effective potential structure at $eB=0.8~{\rm GeV}^2$. 
}
\label{MAMM21_T}
\end{figure}

We next discuss the vacuum stability at finite temperatures, as well as the chiral condensate.
Indeed, for $0<\bar v\leq 2.0\,{\rm GeV}^{-3}$, the vacuum keeps in a stable structure even at high finite temperatures.
Figure~\ref{MAMM21_T} shows the thermal effect on the chiral condensate for $\bar v=2.0{\rm GeV}^{-3}$ and the deformation of the corresponding effective potential. From the panel (b) of Fig.~\ref{MAMM21_T}, one can find that  
the magnetic-dependent AMM hardly affects 
the deformation of the effective potential and 
the vacuum structure is still steady at the thermomagnetic system. 
In this case, looking at the magnetic dependence on the chiral condensate at finite temperatures, one can also find that the enhancement of the chiral symmetry breaking under  the magnetic field is inhibited by the presence of the magnetic-dependent AMM, with keeping the stable vacuum structure.

\begin{figure}[H] 
\begin{tabular}{cc}
\begin{minipage}{0.5\hsize}
\begin{center}
    \includegraphics[width=8cm]{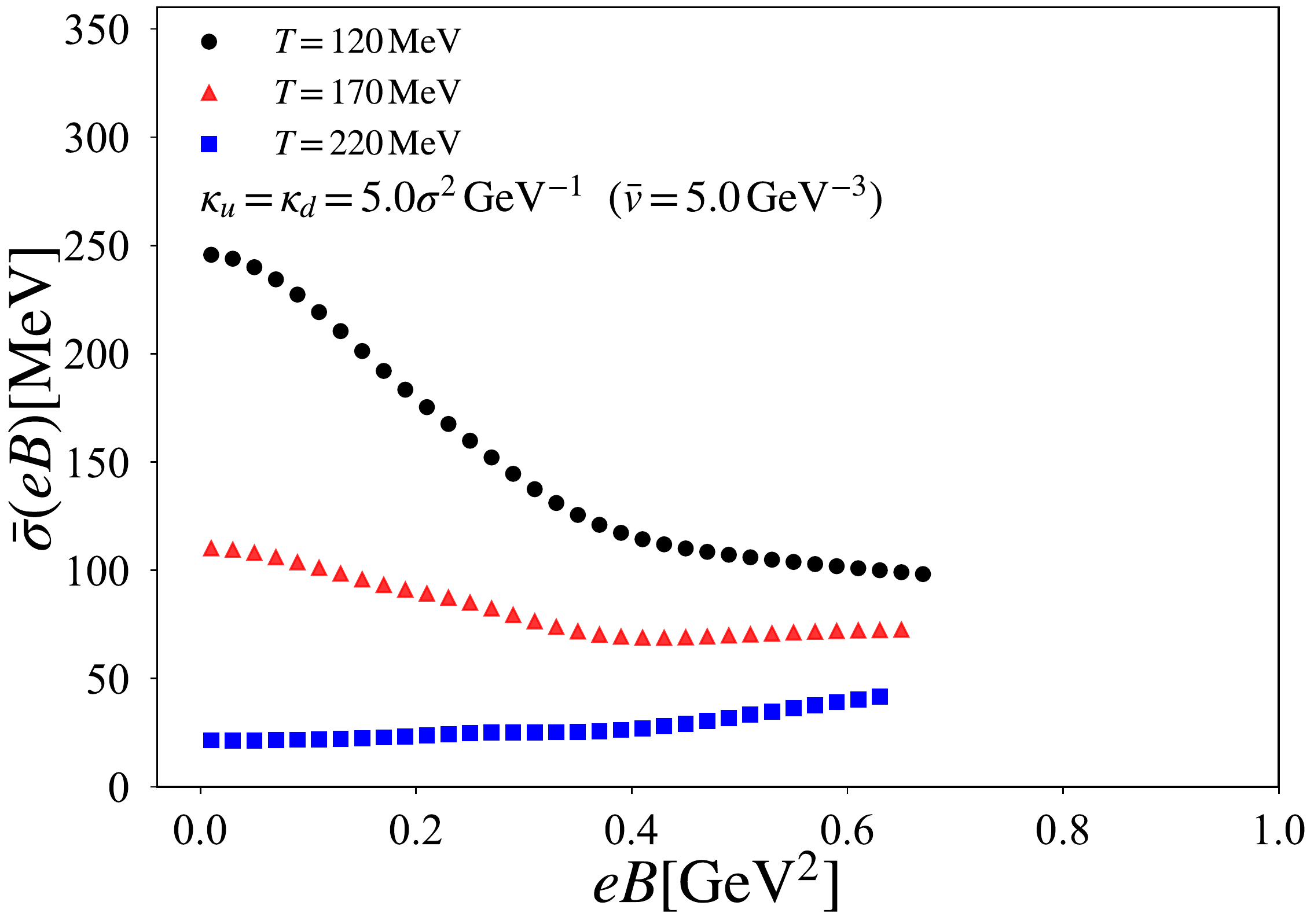}
    \subfigure{(a)}
\end{center}
\end{minipage}
\begin{minipage}{0.5\hsize}
\begin{center}
\includegraphics[width=8cm]{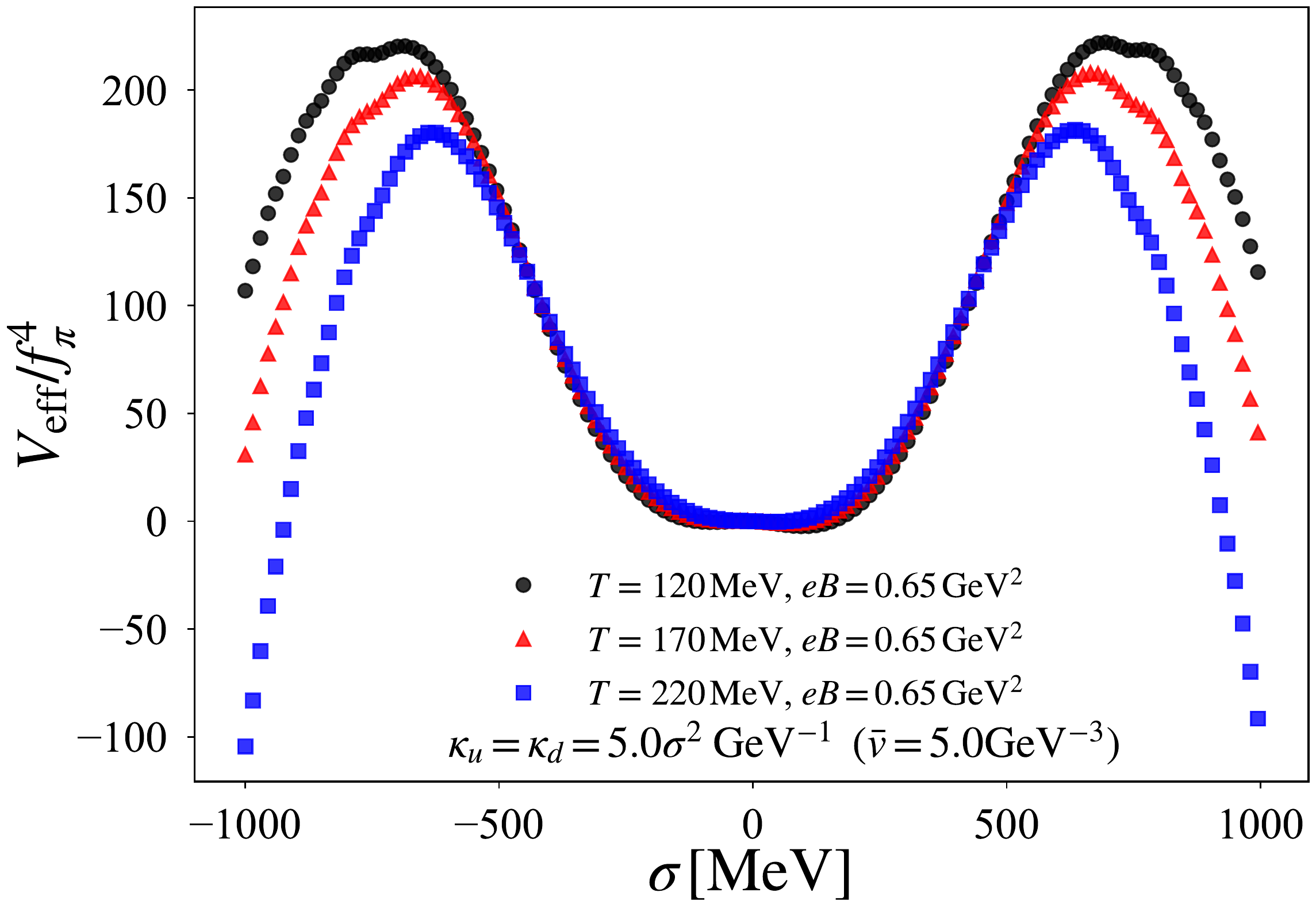}
    \subfigure{(b)}
\end{center}
\end{minipage}
\end{tabular}
\caption{
The thermal effect on the vacuum instability for $\bar v=5.0{\rm GeV}^{-3}$. 
The panel (a): the temperature evolution of the chiral condensate. 
The panel (b): the thermal deformation of the normalized effective potential at $eB=0.65~{\rm GeV}^2$.
}
\label{MAMM25_T}
\end{figure}

For $\bar v>2{\rm GeV}^{-3}$, the thermal effect promotes the destruction of the effective potential structure. 
In Fig~\ref{MAMM25_T}. we take $\bar v=5.0{\rm GeV}^{-3}$ as a case of the large AMM parameter. Actually,
the thermal evolution makes the vacuum structure more unstable, as drawn in the panel (b) of Fig.~\ref{MAMM25_T}.
%
As the consequence of the collapse of the vacuum, we find that the chiral condensate vanishes 
for $eB>0.65~{\rm GeV}^{2}$
(see the panel(a) of Fig.~\ref{MAMM25_T}). 

In the case of the magnetic-dependent AMM proportional ($\kappa_{u,d}=\bar v \sigma^2$) 
with the parameter constraint
$0<\bar v\leq2.0{\rm GeV}^{-3}$, the effective potential keeps in the stable structure while
the unexpected behavior does not emerge in the chiral condensate, such as the induced-first order phase transition and the flip on the sign of $\bar \sigma$. 
Rather than that, the magnetic-dependent AMM sufficiently suppresses the chiral symmetry breaking under the magnetic field and would be expected to reconcile the NJL results with the inverse magnetic catalysis observed in the lattice simulations.
 


\subsection{Subtracted quark condensate}
In this subsection we compare the results of quark condensate obtained from the NJL model by taking into account the magnetic-dependent AMM ($\kappa_u=\kappa_d=\bar v \sigma^2$)
with those from the lattice QCD calculation~\cite{Ding:2020hxw, Ding:2022tqn},

The quark condensate (chiral condensate) involves the ultraviolet divergence at the zero-temperature part and should be renormalized to be a finite quantity.
To eliminate the divergence arising in the quark condensate under the magnetic field,
we use the following  dimensionless quantity of the subtracted quark condensate, 
\begin{eqnarray}
\frac{\Sigma_u+\Sigma_d}{2}=
1-\frac{m_l}{m_\pi^2 f_\pi^2}\left[
\bigl(
\langle \bar uu\rangle(B,T)-\langle \bar uu\rangle(0,T)\bigl)
+
\bigl(
\langle \bar dd\rangle(B,T)
-\langle \bar dd\rangle(0,T)
\bigl)
\right].
\label{subcon}
\end{eqnarray}

\begin{figure}[H] 
\begin{tabular}{cc}
\begin{minipage}{1\hsize}
\begin{center}
    \includegraphics[width=9cm]{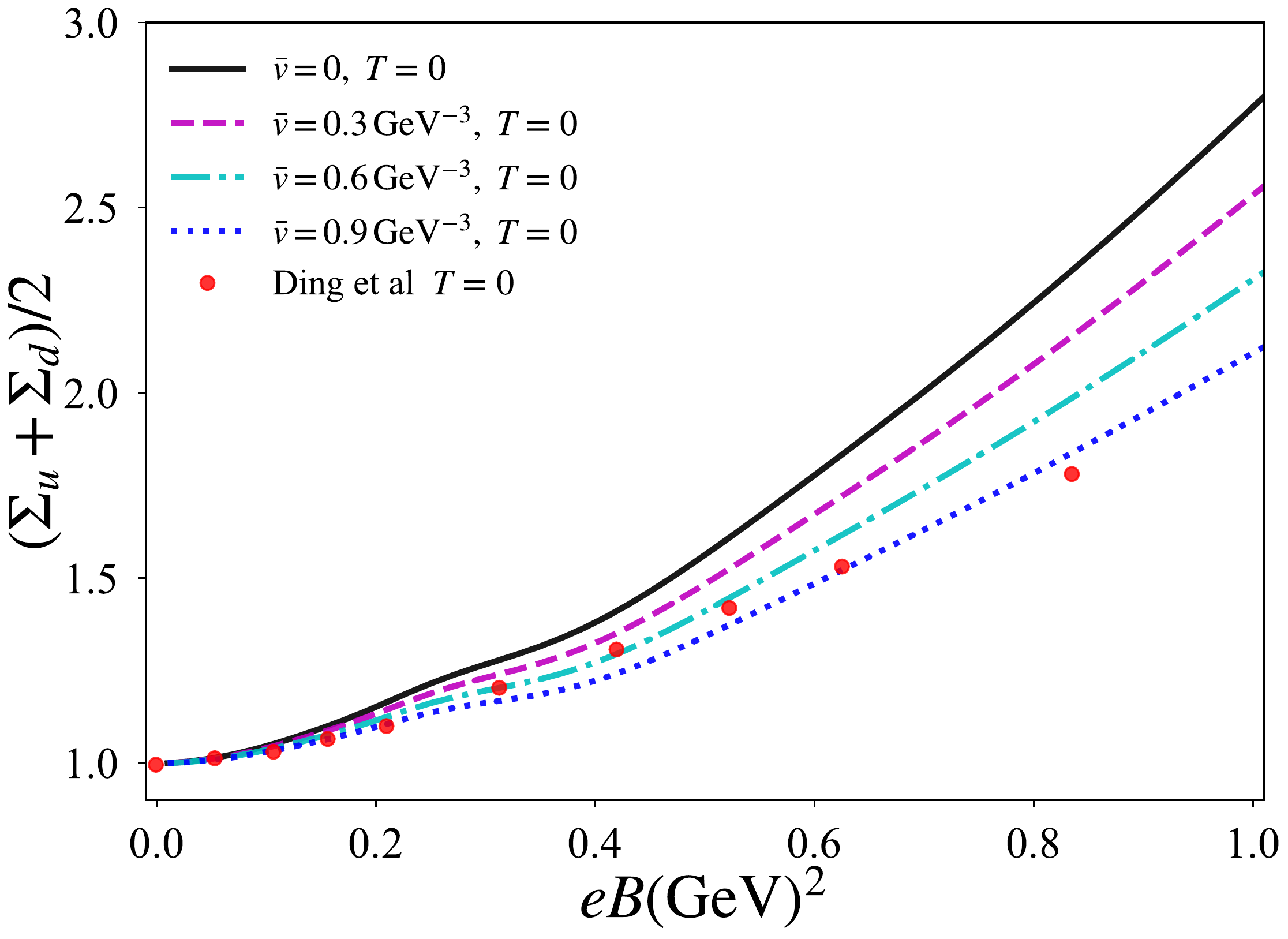}
\end{center}
\end{minipage}
\end{tabular}
\caption{
Comparison of the subtracted quark condensate
between the NJL result and 
the lattice QCD data
at $T=0$ \cite{Ding:2020hxw}.
}
\label{sub_con_T0}
\end{figure}

Figure~\ref{sub_con_T0} shows the magnetic dependence on the subtracted quark condensate at zero temperature,   
in comparison with the lattice QCD observation~\cite{Ding:2020hxw}.
The subtracted quark condensate, in the case
of the NJL model without the AMM term ($\kappa_{u,d}=0$), monotonically grows as the magnetic field increases, i.e., the magnetic catalysis effect. 
This predicted magnetic dependence deviates from the lattice QCD data.
The NJL result in the absence of the AMM contribution tends to predict a larger chiral symmetry breaking under the magnetic field.
However, by including the magnetic-dependent AMM ($\kappa_u=\kappa_d=\bar v \sigma^2$), 
the magnetic enhancement in the subtracted quark condensate is suppressed and the the NJL result 
for $\bar v=0.9\,{\rm GeV}^{-3}$  is in good agreement with the lattice QCD results. 


\begin{figure}[H] 
\begin{tabular}{cc}
\begin{minipage}{0.5\hsize}
\begin{center}
    \includegraphics[width=8cm]{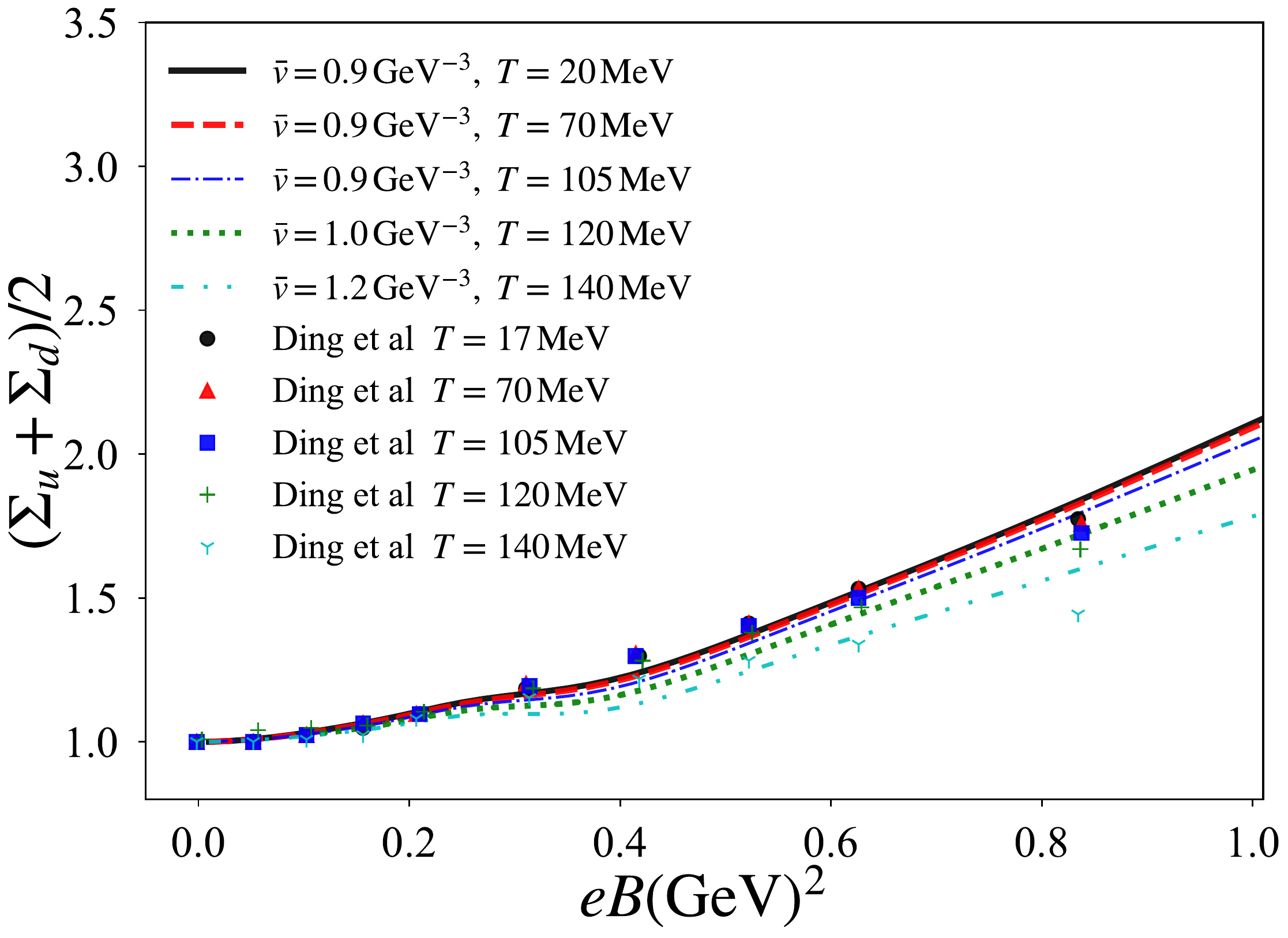}
    \subfigure{(a)}
\end{center}
\end{minipage}
\begin{minipage}{0.5\hsize}
\begin{center}
    \includegraphics[width=8cm]{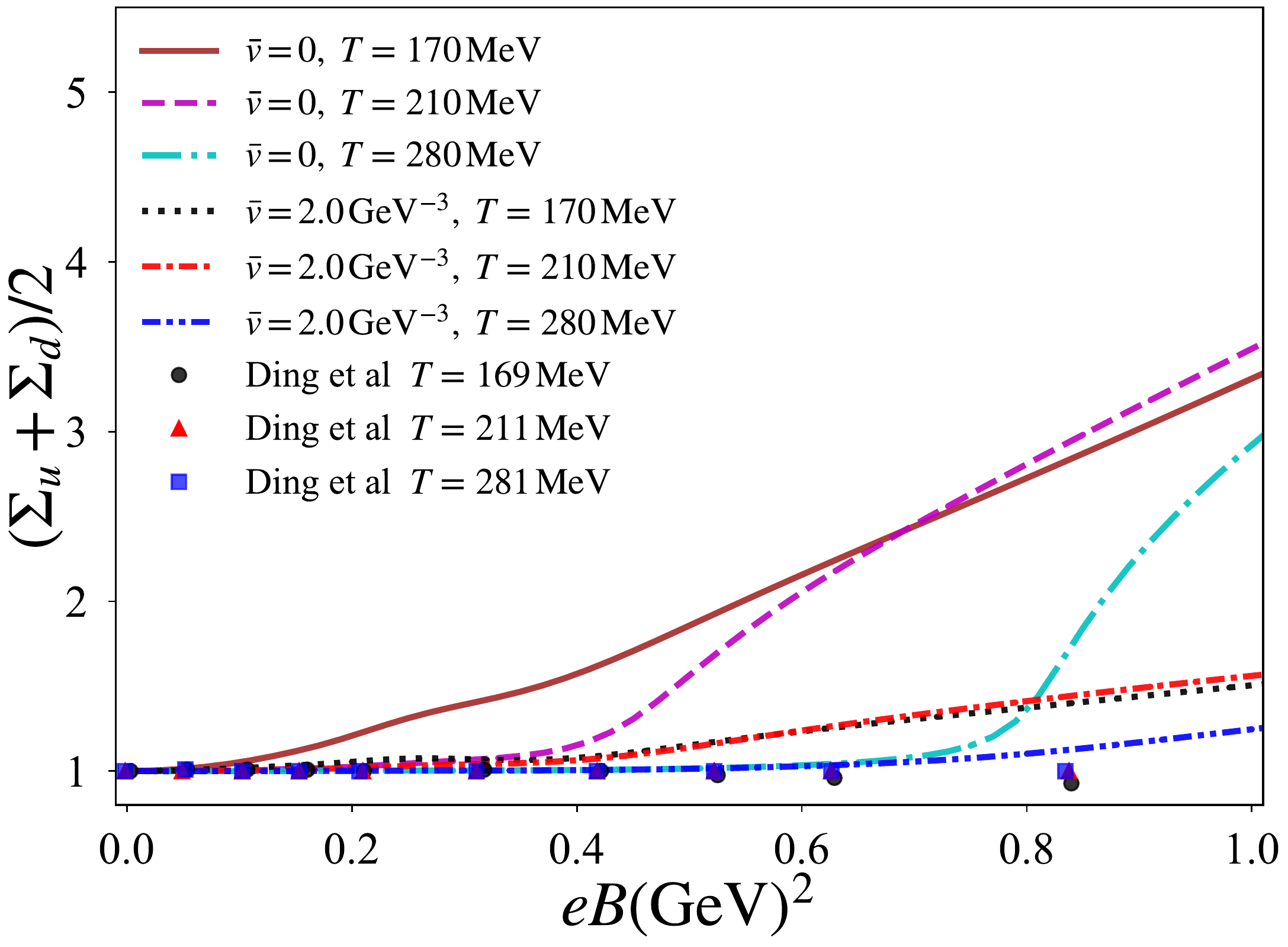}
    \subfigure{(b)}
\end{center}
\end{minipage}
\end{tabular}
\caption{
Comparison of the subtracted quark condensate between the NJL result and the lattice QCD data \cite{Ding:2022tqn} for 
(a): the low temperature regions ($T\sim0- 140~{\rm MeV}$), and
(b): the high temperature regions ($T\sim 170-280~{\rm MeV}$).
}
\label{sub_con_Tfinite}
\end{figure}

 Figure~\ref{sub_con_Tfinite} displays the comparison of the subtracted quark condensate between the NJL model prediction and the lattice QCD data at finite temperatures \cite{ Ding:2022tqn}.
To reconcile the gap between the NJL result and the lattice result, we tune the value of the AMM parameter $\bar v$. 
For the low temperature regions where $T\sim0- 140~{\rm MeV}$, 
the panel (a) of Fig.~\ref{sub_con_Tfinite} shows that the NJL results fit well with the lattice QCD. 
On the other hand, 
for the high temperature regions where $T\sim 170-280~{\rm MeV}$,
the AMM term also reduces the magnetic enhancement of the quark condensate, but the NJL results start to deviate from the lattice observation at around $eB=0.4\,{\rm GeV}^2$ for $T=170$ MeV and $T=210$ MeV, and at around $eB=0.6\,{\rm GeV}^2$ for $T=280$ MeV,
as shown in the panel (b) of Fig.~\ref{sub_con_Tfinite}.
Due to the restriction of the AMM parameter ($0\leq \bar v\leq  2\,{\rm GeV}^{-3}$), the NJL results can not perfectly fit the lattice observation in whole magnetic field regions. 
Although the deviation remains at around the high temperature regions ($T\sim 170-280~{\rm MeV}$), 
the magnetic dependence of the subtracted quark condensate
is sufficiently suppressed by the magnetic-dependent AMM and 
qualitatively agrees with the lattice data.




\begin{figure}[H] 
\begin{tabular}{cc}
\begin{minipage}{1\hsize}
\begin{center}
    \includegraphics[width=9cm]{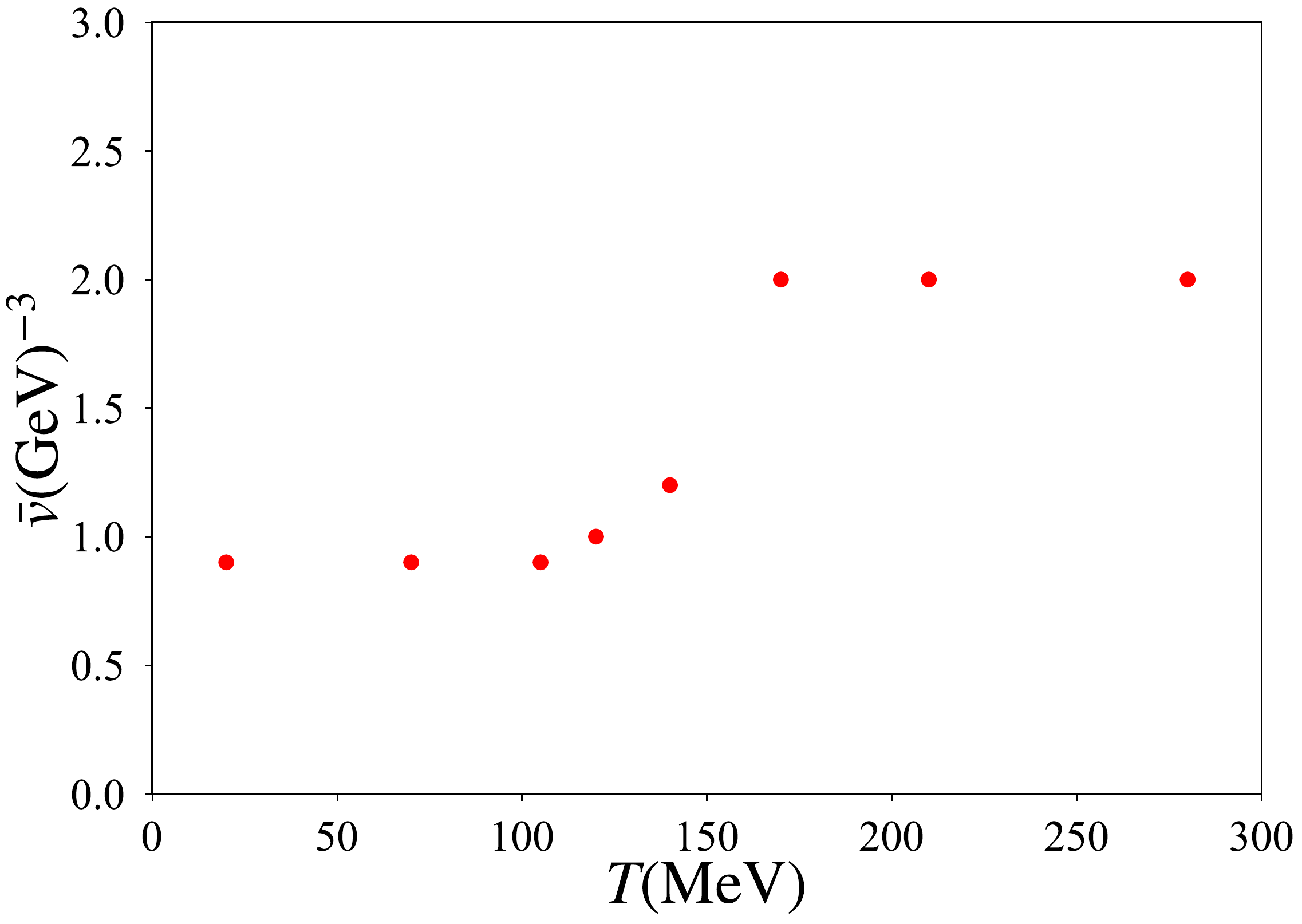}
\end{center}
\end{minipage}
\end{tabular}
\caption{
Temperature dependence on the AMM parameter $\bar v$. 
}
\label{ITD_barv}
\end{figure}

With the comparisons of the subtracted quark condensate, we discuss the temperature dependence of the AMM term.
The intrinsic temperature dependence of the AMM parameter $\bar v$ can be read out from Figs.~\ref{sub_con_T0} and~\ref{sub_con_Tfinite}, and is plotted in 
Fig.~\ref{ITD_barv}. Around low temperature regions where $T\sim0-100$ MeV, the AMM parameter $\bar v$ behaves as a constant. As the temperature further increases,  
$\bar v$ rapidly grows up at around the pseudocritical temperature  
$T_{\rm pc}\simeq 160$ MeV ($T_{\rm pc}$ will be seen later)
and then reaches at the upper limit of $\bar v$, $\bar v=  2\,{\rm GeV}^{-3}$.  
This thermal behavior would imply that 
the intrinsic temperature dependence on $\bar v$ may correlate with the chiral phase transition.

\begin{figure}[H] 
\begin{tabular}{cc}
\begin{minipage}{1\hsize}
\begin{center}
    \includegraphics[width=8cm]{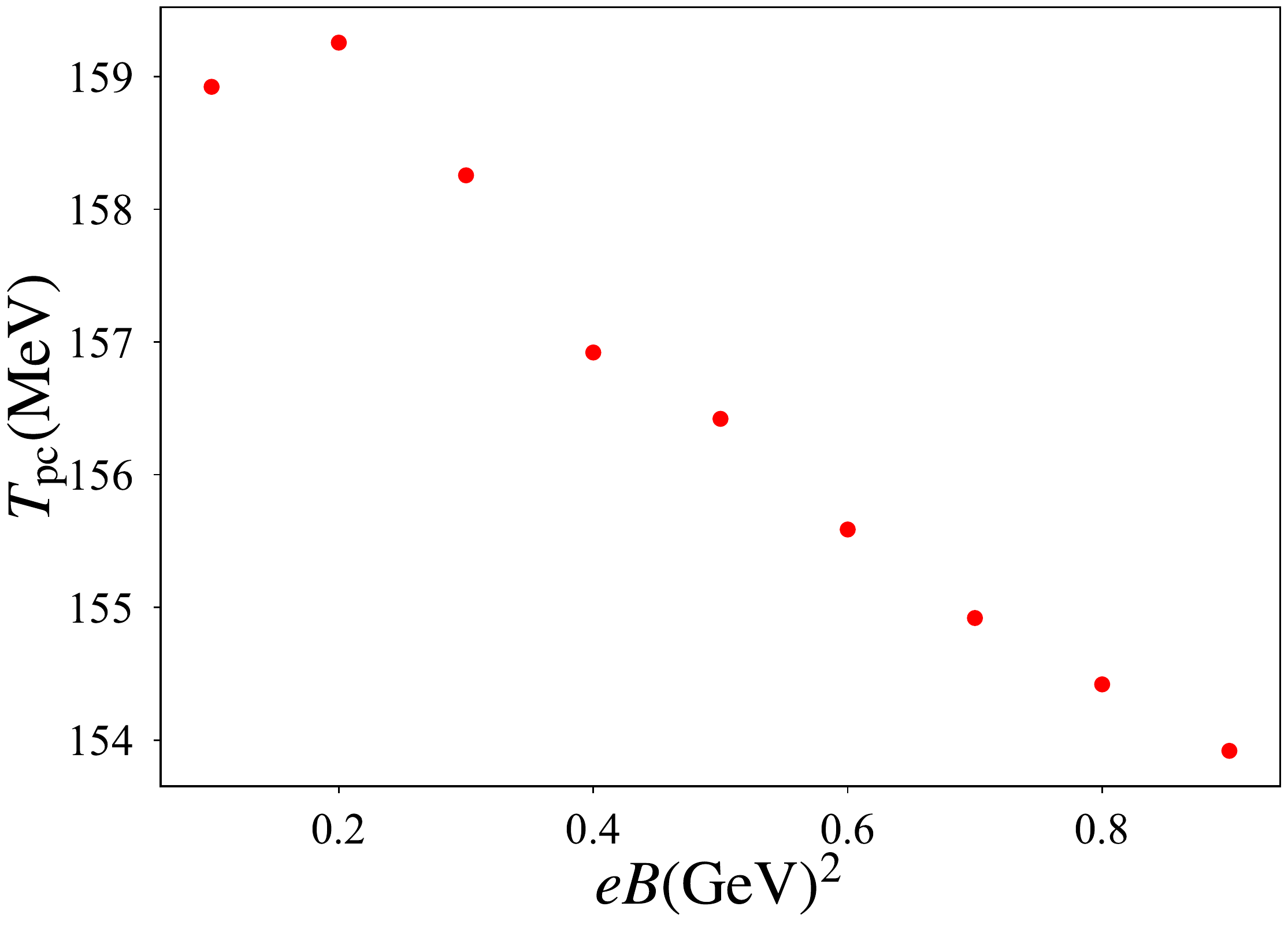}
\end{center}
\end{minipage}
\end{tabular}
\caption{
Chiral phase diagram under the constant magnetic field
based on the NJL model with the fitting-parameter $\bar v$.
}
\label{PD_eB}
\end{figure}

Using the temperature-dependent AMM parameter $\bar v(T)$ estimated in Fig.~\ref{ITD_barv},
we finally show the chiral phase diagram for the $eB-T$ plane
in Fig.~\ref{PD_eB}
\footnote{
To evaluate the psuedocrtical temperature from the inflection point of the quark condensate with respect to temperature, we
have used the alternative expression of the subtracted quark condensate,
\begin{eqnarray*}
\frac{\bar\Sigma_u+\bar\Sigma_d}{2}=
1-\frac{m_l}{m_\pi^2 f_\pi^2}\left[
\bigl(
\langle \bar uu\rangle(B,T)-\langle \bar uu\rangle(0,0)\bigl)
+
\bigl(
\langle \bar dd\rangle(B,T)
-\langle \bar dd\rangle(0,0)
\bigl)
\right].
\label{subcon}
\end{eqnarray*}
}.
This figure shows that the pseudocritical temperature decreases with the development of the magnetic field, and  the inverse magnetic catalysis is surely provided by the NJL analysis with the parameter $\bar v(T)$ (fitted by using the lattice observation \cite{Ding:2022tqn})
\footnote{  
The tendency of the inverse magnetic catalysis 
is more prominent in the other lattice observation \cite{Bali:2012zg} where 
the physical pion mass is taken and the results is  extrapolated to the continuum limit.
}
This indicates that the magnetic-dependent AMM with the intrinsic temperature dependence
($\kappa_{u,d}=\bar v(T)\sigma^2$) somewhat inhibits the magnetic catalysis.









\section{Summary and discussion}
\label{sec3}
In this paper, we have explored 
the effective form of the AMM in the thermomagnetic QCD vacuum.
Employing the NJL model with the effective interaction of the AMM for quarks, 
we discussed the influence of the AMM on the chiral condensate in the following three forms separately:
(a) $\kappa_{u,d}={\rm const.}$,
(b) $\kappa_{u,d}=v \sigma$ and
(c) $\kappa_{u,d}=\bar v \sigma^2$.
It was shown that 
the AMM interaction drastically affects the chiral phase transition as well as the effective potential structure.
What we have found in the three forms can be summarized as follows:
\begin{itemize}
\item[(a)]
For $\kappa_{u,d}={\rm const.}$,
the constant AMM significantly deforms the effective potential structure. 
The potential wall is driven by the constant AMM, so that the chiral first order phase transition is induced at zero temperature and finite temperatures.
However, this induced-first phase transition is inconsistent with the lattice QCD result which shows a crossover for chiral phase transition.  

\item[(b)]
For $\kappa_{u,d}=v \sigma$, 
the effective potential structure is also deformed by
the magnetic-dependent AMM. 
As a result, the global minimum point of the effective potential jumps from the positive vacuum to the negative vacuum at a critical magnetic field. Thus, the sign of the chiral condensate turns to be flipped from positive to negative with the increase of the magnetic field at $T=0$ and $T\neq 0$. However the flip on the sign of the chiral condensate (subtracted quark condensate) has not been observed in the lattice QCD simulations. 

\item[(c)]
For $\kappa_{u,d}=\bar v \sigma^2$, the accidental jumps do not emerge in the chiral condensate across the chiral phase transition, such as the induced-first order phase transition and the sign flip on the chiral condensate.
Rather than that, 
the magnetic-dependent AMM plays a role of inhibiting  
the chiral symmetry breaking under the magnetic field. 
However, the instability happens in the effective potential for the large value of the AMM parameter $\bar v$. Hence, the AMM parameter $\bar v$ should be restricted to provide the inhibition for the magnetic catalysis with keeping the stable vacuum.
Actually, 
by using the restricted AMM parameter $\bar v$, the subtracted quark condensate in the NJL model  becomes not only qualitatively but also quantitatively in good agreement with the lattice results at the zero temperature and finite temperatures.
\end{itemize}

Our findings indicate that the magnetic-dependent AMM form $\kappa_{u,d}=\bar v \sigma^2$
would be the practicable effective form to adequately describe the property of the thermomagnetic vacuum of QCD.
In addition, we also provided the intrinsic temperature dependence of the AMM parameter $\bar v$ through the comparison between the NJL results and the lattice QCD results.  This fitting-parameter $\bar v$ implies that 
the temperature dependence on the AMM 
may be correlated to the chiral phase transition.


Before closing this paper, we shall make some comments on the 
implications of the magnetic-dependent AMM.
The recent lattice QCD simulation also exhibits the magnetic effect on the meson masses 
at zero-temperature \cite{Ding:2020jui}.
This shows 
that the pions mass starts to deviate from the piont-like particle behavior with the increase of the magnetic field.
To address this deviation,
exploring the AMM contribution on
the meson properties would be worth studying.
Actually, the influence of 
the magnetic-dependent AMM on the meson masses has already been discussed based on the NJL model \cite{Xu:2020yag,Lin:2022ied}. However, the vacuum structure
and its stability have not been taken into account.
Hence, 
it would be valuable to reevaluate the meson properties with the vacuum structure in mind.

What’s more,
the magnetic susceptibility
at the thermomagnetic QCD vacuum
has recently been  revealed  by the lattice QCD simulation \cite{Bali:2012jv,Bali:2020bcn}: the magnetized QCD vacuum behaves  
the diamagnetism at low temperatures, but 
turns out to be 
the paramagnetism with the increase of the temperature. 
It would be important to 
investigate the influence of the quark AMM on the magnetic susceptibility. Indeed, the NJL model tends to produce the paramagnetism  at finite temperatures, which is enhanced by  
the presence of the magnetic-dependent AMM (without the intrinsic temperature dependence of the AMM parameter)
~\cite{Xu:2020yag,Lin:2022ied}.
It would be also worth to 
consider the intrinsic temperature dependence on the magnetic susceptibility.


As an  application, 
the predicted magnetic-dependent  AMM interaction would be applied to the high-dense matter physics with the magnetic field.
In particular,
the strong magnetic field is generated in neutron stars or magnetars.
The mechanism of the 
the generation of the strong magnetic field has not been clarified yet.
In Ref.~\cite{Tatsumi:1999ab,Maruyama:2000cw}, it was pointed out that the spin polarization of baryons induces the ferromagnetism in magnetars.
The quark-spin 
is crucially related to the spontaneous magnetization. 
In addition, 
it has been discussed that
the spontaneous magnetization is driven by the quark AMM  
in the NJL model having the constant AMM term \cite{Tsue:2015lea}.
Therefore, we anticipate that 
the spontaneous magnetization can be  directly linked with the dynamical chiral symmetry breaking by using the magnetic-dependent AMM instead of constant one. 
The magnetic-dependent AMM would  be a significant ingredient to understand 
a new aspect of magnetized quark matters as well as magnetars.


In this study, 
we have supposed that 
the chiral condensate and the AMM take the isospin symmetric form: $\bar \sigma_u=\bar \sigma_d$ and $\kappa_u=\kappa_d$. In fact, the chiral symmetry is explicitly broken by the external magnetic field, so that the flavor symmetry breaking should be taken into account in the the chiral condensate and the AMM term. We leave the flavor symmetry breaking effect to further study.






\begin{acknowledgements}
We are grateful to discussions with Fan Lin, Kun Xu and Aminul Chowdhury.
This work is supported in part by the National Natural Science Foundation of China (NSFC) Grant  Nos 11725523, 11735007, and supported by the Strategic Priority Research Program of Chinese Academy of Sciences under Grant Nos XDB34030000 and XDPB15, the start-up funding from University of Chinese Academy of Sciences(UCAS), and the Fundamental Research Funds for the Central Universities.
\end{acknowledgements}


\end{document}